\documentclass[ %reprint,
superscriptaddress,
preprint,
showpacs,preprintnumbers,
 amsmath,amssymb,
aps,
pra,
]{revtex4-1}

\usepackage{graphicx}% Include figure files
\usepackage{dcolumn}% Align table columns on decimal point
\usepackage{braket}
\usepackage{bm}% bold math
\usepackage{color}
\begin{document}

%\preprint{APS/EIT-EIA-AT}

\title{Interplay between electromagnetically induced transparency (EIT), absorption (EIA) and Autler-Townes (AT) splitting in $\mathcal{N}$-type atomic system: An experiment and theory}.% Force line breaks with \\
\author{Bankim Chandra Das}
\email{bankim.das@saha.ac.in}
\affiliation{Saha Institute of Nuclear Physics, HBNI, 1/AF, Bidhannagar,
	Kolkata -- 700064, India.}
\author{Arpita Das}
\affiliation{Saha Institute of Nuclear Physics, HBNI, 1/AF, Bidhannagar,
	Kolkata -- 700064, India.}
\author {Dipankar Bhattacharyya}
%\email{bh.dipankar@gmail.com} 
\affiliation{Department of Physics, Santipur College, Santipur, Nadia, West Bengal, 741404, India.}
\author{Shrabana Chakrabarti}
\affiliation{Saha Institute of Nuclear Physics, HBNI, 1/AF, Bidhannagar,
	Kolkata -- 700064, India.}
\author{Sankar De}
\email{sankar.de@saha.ac.in}
\affiliation{Saha Institute of Nuclear Physics, HBNI, 1/AF, Bidhannagar,
	Kolkata -- 700064, India.}

\begin{abstract}
In this article we have shown that the atomic states can be engineered by tunning the coupling Rabi frequency for a system with $\mathcal{N}$-type configuration. Electromagnetically induced transparency (EIT), Electromagnetically induced absorption (EIA) and Autler-Townes (AT) splitting has been observed experimentally in a four level $\mathcal{N}$-type atomic vapor of $^{85}Rb$ atoms in the hyperfine levels of $D_2$ transition. It has been shown that the response of the atomic medium can be tunned from highly transparent to highly absorptive in our case. The evolution of the atomic states from the dark state $\ket{D}$ to the non-coupled state $\ket{NC}$ has been studied with the partial dressed state approach which makes the backbone of the modification of the atomic response. In addition, transient solutions in the time domain and steady state solution in the frequency domain has been studied. The population dynamics and the coherence contribution in each case has been analyzed by the time dependent solutions. The experimentally observed steady line-shape profiles has been supported by the steady state solution of optical-Bloch equations considering the Maxwell-Boltzmann velocity distributions of the atoms. It has been observed that the crossover between the EIT and the AT splitting has been replaced by the interference contribution of the EIA in this $\mathcal{N}$-type system.   
\end{abstract}

\maketitle
\section{Introduction}
Atomic coherence is a kind of knob to control the medium response coherently. Electromagnetically induced transparency (EIT) \cite{kasapi95} is such a gate that can be created in a three level atom interacting with two beams. Due to the coherent interaction between the atom and the field, many technological advances such as slowing of light pulses \cite{Hau99}, storing of light \cite{lukin2001}, magnetometer \cite{shah2007}, white light cavity \cite{WICHT1997}, lasing without inversion (LWI) \cite{zibrov1995,scully1992} etc. have been achieved recently. 

More complicated systems can be formed which exhibits EIT, if another field is added\cite{Scully2009}. Three beam spectroscopy can be more interesting than the former because it is preferable in order to study the non-linear properties of the atomic medium \cite{krmpot}. Like the EIT, another coherent phenomena is electromagnetically induced absorption (EIA) \cite{lezama,BankimJCP,charles} which can be observed with three beam spectroscopy. Due to EIA, the system becomes highly absorptive and sharp spectral features can be observed,  which have several applications in precision spectroscopy. Lezama et al. \cite{lezama} had shown that a minimum four level atomic system is necessary to achieve the EIA. But in a $\Lambda$ type system if the pump field is made to be the standing wave, EIA can be observed \cite{bae,chanu2012}. Subluminal to superluminal light can be observed in this system \cite{Bae10}.

Three beam spectroscopy with four level atom is preferable in order to observe the EIA. There are a few studies where four level atoms interacting with the three beams are studied. But most of the work on the four level system has concentrated on the three ground states and one excited states {\cite{Goren_tripod2004,Ham2000,wang2017strong}}. Such systems have shown to exhibit the enhanced Kerr non-linearities \cite{Niu2005}. Another type of four level system is a $\mathcal{N}$-type system with two ground states and two excited states \cite{Goren2004, mahapatro2009,KONG2007}. This system is of particular interest due to the strong Kerr effect\cite{schmidt96}. In this system EIA can occur. Theoretical \cite{scully1999} and experimental studies \cite{lukin2003,mahapatro2009} have shown that due to the double dark states, the transformation of EIT to EIA can be observed. There is two types of $\mathcal{N}$-type system that can be formed. One which starts with the weak probe beam, followed by the strong coupling and pump beams \cite{mahapatro2009}. The other one where the probe lies in between the pump and the coupling fields \cite{KONG2007}. However most of the studies in the $\mathcal{N}$-type system consist of EIT and EIA, this system can also exhibit Autler-Townes (AT) \cite{autler1955} effect. AT splitting occurs when the atomic spectral line splits into two absorption lines having a transparency between them. Here the Fano resonances \cite{Fano61} are not observed whereas in EIT, the Fano interference occurs. There are previous studies where the transformation from EIT to EIA have been seen \cite{bae,chanu2012}. Studies also show the transformation from EIT to AT \cite{barry2011}. Recent studies show the transformation of EIA to AT \cite{Islam2017} and EIA to EIT \cite{kim2001}. The $\mathcal{N}$-type system opens up the pathway to study the interplay between the EIT, EIA and AT effects in a single atomic system.

In this paper we have studied both experimentally and theoretically the four level $\mathcal{N}$-type system with three beams in $D_2$ transition of $^{85}Rb$ in a closed system configuration. Here we have observed the EIT, EIA and the AT splitting depending only on the pump and the coupling Rabi frequencies. We have scanned the pump beam and observed the probe transmission. The coupling Rabi frequency ($\Omega_{b}$) was varied taking the pump Rabi frequency ($\Omega_{c}$) as a parameter. Depending on the coupling Rabi frequency $\Omega_{b}$, the transformation from EIT to EIA and from EIA to AT was studied. In order to explain the underlying physical phenomena we have solved the optical Bloch equations (OBE) for a four level $\mathcal{N}$-type system interacting with three optical fields. We have performed the time dependent solutions in order to understand the population dynamics of the different states and the coherence contributions. We also studied the steady state solution considering the thermal velocity of the atoms. We further explained the observed phenomena with the partial dressed state \cite{nakajima1999} approach. 
\section{Experimental setup}

\begin{figure}[h]
\centering
\includegraphics[scale=.32]{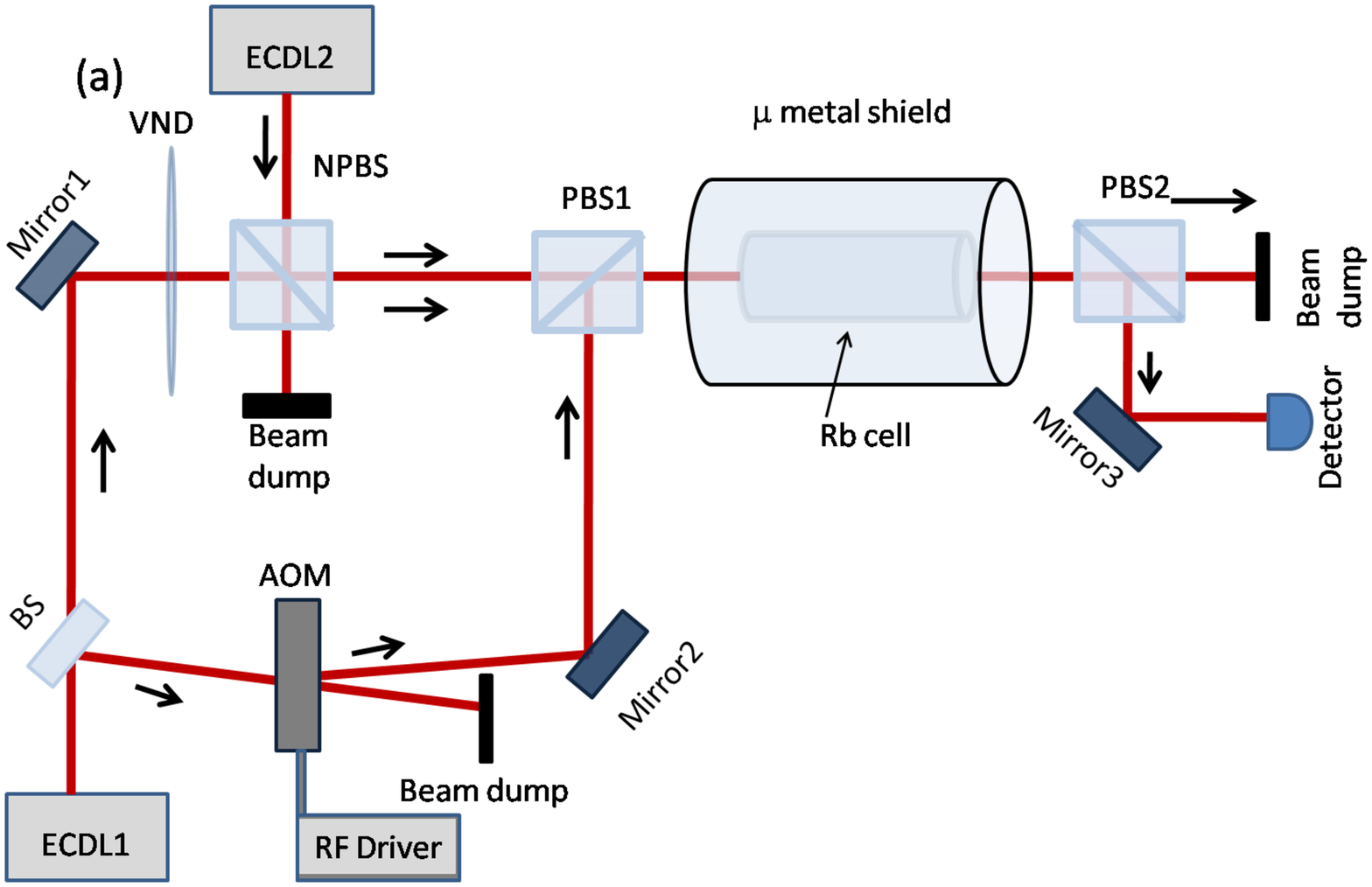}
\includegraphics[scale=.42]{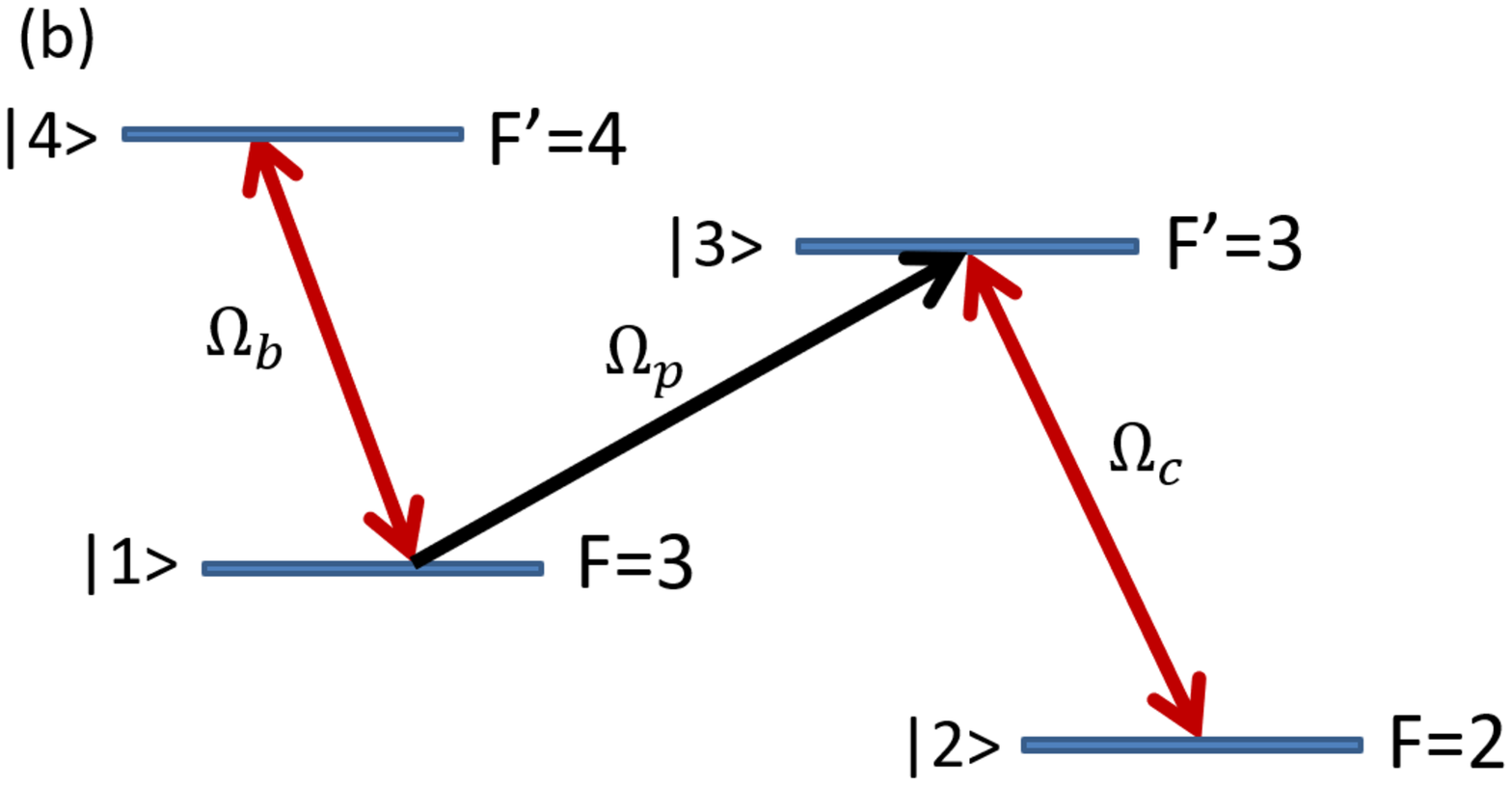}

\caption{(a) Experimental setup to study the N-type system. ECDL: External cavity diode laser, BS: beam splitter, VND: variable neutral density filter, NPBS: Non polarizing cubic beam splitter, PBS: polarizing cubic beam splitter, AOM: acousto optic modulator, Rb Cell: Rubidium vapour cell. (b) Energy level diagram for $^{85}Rb$ $D_2$ transition. $\Omega_{p}$, $\Omega_{b}$, and $\Omega_{c}$ are the probe, the coupling and the pump Rabi frequencies respectively.   }
\label{Eperimental setup}
\end{figure}

The experimental setup to study the coherent phenomena of the $\mathcal{N}$-type system in the hyperfine levels of Rb atoms is shown in figure \ref{Eperimental setup}. To make the $\mathcal{N}$-type system we need three laser beams. We have used two different external cavity diode lasers (ECDL) in order to generate the three beams namely, the coupling beam, the pump beam and the probe beam. The ECDLs were lasing at 780 nm (DL100 Toptica) and having a beam diameter $\sim 2mm$ and a line-width $\sim 1 MHz$.
 
The coupling beam was taken from ECDL1 and the probe beam was generated by down-shifting the coupling beam by $120$ $MHz$ with the help of an acousto optic modulator (AOM). From ECDL1 the output laser beam was incident on a $30:70$ plate beam splitter (BS). The transmitted part from BS was reflected by mirror1 and was passed through a $50: 50$ non polarizing cubic beam splitter (NPBS). The transmitted part was again passed through a polarizing beam splitter cube (PBS1). The transmitted part of this PBS1 was p-polarized and it was used as the coupling beam. The coupling beam was passed through the Rb cell. The reflected part from BS of the initial laser beam was passed through the AOM. We took the $-1^{th}$ diffracted beam as the probe beam. The probe beam was reflected by another mirror (Mirror2) and incident on the reflecting surface of the PBS1. The probe beam was mixed with the coupling beam in PBS1 and passed through the Rb cell co-propagating and co-linearly. A small part of the output of ECDL1 was taken to the saturation absorption setup (not shown in the figure \ref{Eperimental setup}) in order to lock the laser frequency. The locking was done with the help of a lock in amplified (LIR) and a proportional integrator differentiator (PID) loop.

The pump beam is taken from ECDL2. It was incident on the reflecting surface of NPBS and mixed with the coupling beam. Now both the coupling beam and the pump beams were transmitted through PBS1 where both were chosen to be p-polarized. Both these beams and the probe beam were passed through the Rb cell co-propagating and co-linearly. The probe beam was separated from the pump and the coupling beams after the Rb cell with the help of another PBS2. The reflected part from PBS2 was detected with a pin photo-diode (Newport, Model: 2001 ) which gave us the probe transmission spectra. 

We have chosen $^{85}Rb$-$D_2$ transition in order to make the $\mathcal{N}$-type system. The coupling beam was locked to the closed transition $F=3 (\ket{1}) \rightarrow F'=4 (\ket{4})$ of $^{85}Rb$ $D_2$ transition. The probe beam was down-shifted by $120$ $MHz$ with help of the AOM so that it can be locked to the transition $F=3 (\ket{1}) \rightarrow F'=3 (\ket{3})$ (see figure \ref{Eperimental setup}(b)). In this way a V-type system was formed maintaining the phase coherency. In order to make the N-type system the pump beam was introduced in between the transition $F=2 (\ket{2}) \rightarrow F'=3 (\ket{3})$. The pump beam was taken from separate laser ECDL2. For our experiment we have scanned the pump beam and locked the probe and the coupling beams in order to get rid of the Doppler background \cite{BankimJCP,dipankar}. The pump beam made a $\Lambda$-type system with the probe beam whereas the coupling beam made a V-type system with the probe beam. As a whole this $\mathcal{N}$-type system can also be thought of a combination of $\Lambda $ + V-type systems. For our experiment a $50mm$ long and $25mm$ diameter cylindrical cell was used containing both $^{ 85}Rb$ and $^{87}Rb$ in their natural abundances with no buffer gas. The pressure of the cell was $ 10^{-7}$ Torr in the room temperature. The Rb cell was put inside a $\mu$-metal shield. To change the pump intensity (Rabi frequency $\Omega_{c}$) we have used the plate neutral density filters and for changing the coupling beam intensity (Rabi frequency $\Omega_{b}$) a variable ND filter (VND) was used as shown in the figure \ref{Eperimental setup}(a). Rabi frequency ($\Omega$) was calculated from the intensity ($I$) of the laser beam using the relation $\Omega = \Gamma\sqrt{\frac{I}{2I_{sat}}}$ \cite{steck}. $\Gamma$ is the natural line-width and $I_{sat}$ is the saturation intensity.
  
\section{Experimental results}
%In this experiment we choose $^{85}Rb$ $D_{2}$ trasition to form the N-type system. We have scanned the pump beam from $F=2 \rightarrow F'=3$ transtion. We have locked the prob beam in the transition $F=3 \rightarrow F'=3$. In this way a $\Lambda$-type system is formned. To make the N-type system We have locked the coupling beam in the transition $F=3 \rightarrow F'=4$. The probe beam is actually generated from the coupling beam by downshifting it with the help of an AOM by $-120MHz$. So the probe and coupling beam is phaselocked and the pump beam is generated from another independant laser. We have locked the coupling beam and probe beam and scanned the pump beam in order to get rid of the doppler back ground.\\

\begin{figure}[h]
	\centering
\includegraphics[scale=.11]{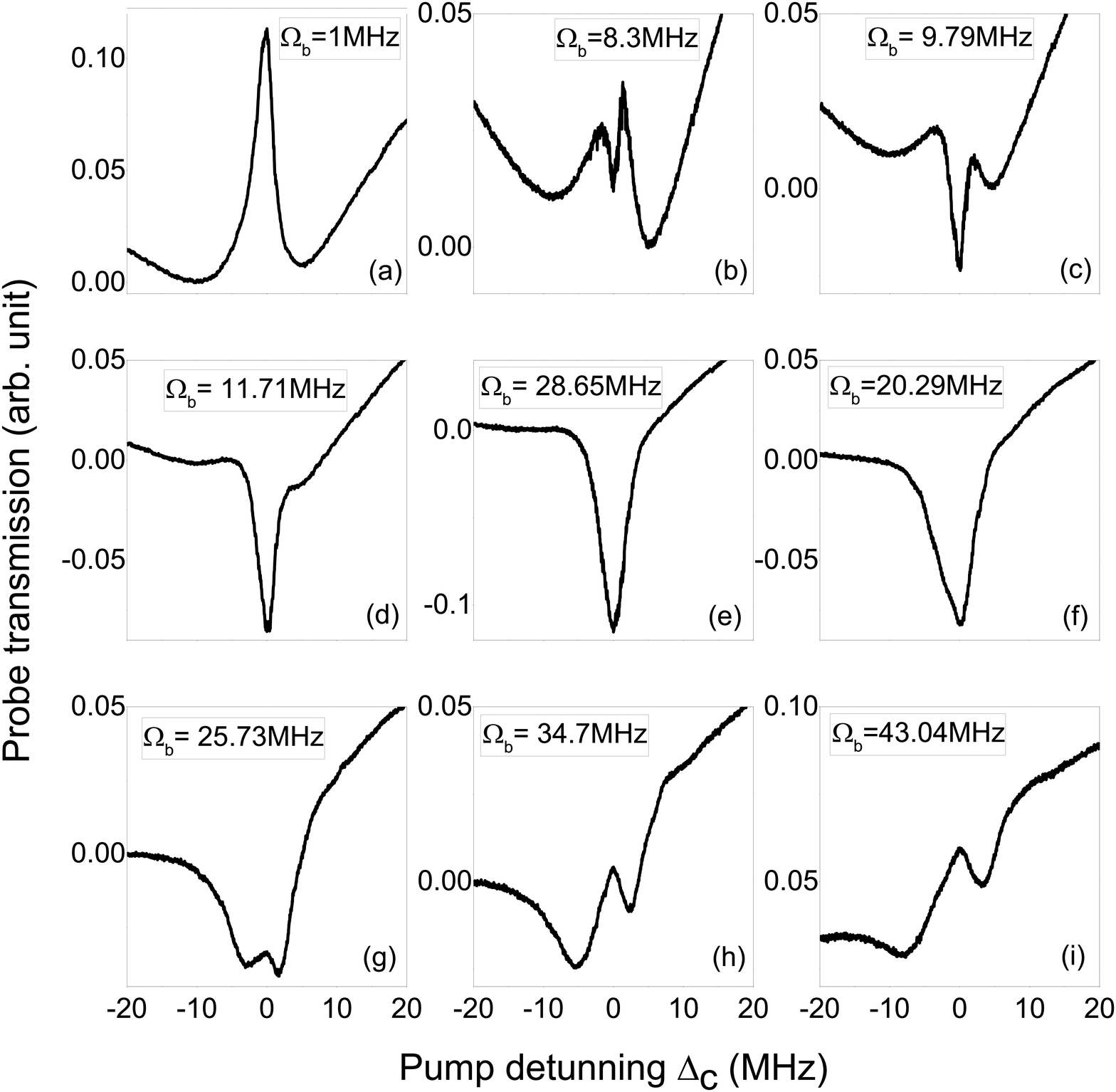}	
	\caption{Experimentally observed probe transmission as a function of pump detunning. The pump Rabi frequency $\Omega_{c}$ was fixed at $17.57$ $MHz$ and the coupling Rabi frequency $\Omega_{b}$ was varied continuously (see legends).}
	\label{pump_power_1.25mw}
\end{figure}

In this experiment our main goal was to show the EIT, EIA and AT splitting in a single system by just tunning the coupling Rabi frequency $\Omega_{b}$. We have studied how the response of probe transmission of $\mathcal{N}$-type system modifies in presence of the coupling beam along with the pump beam. We also wanted to study how the dark states evolved in the presence of the third beam namely, the coupling beam. We have neglected the optical pumping effects of the nearby transitions and focused on the phenomena occurring only in the vicinity of the transition $F=3 \rightarrow F'=3$.

In the experiment the probe Rabi frequency $\Omega_{p}$ was always kept fixed at $2.51$ $MHz$. To study the coupling beam variations we have varied the coupling  Rabi frequency $\Omega_{b}$ from $1$ $MHz$ to $43.04$ $MHz$, taking the pump Rabi frequency $\Omega_{c}$ as a parameter. We took three different pump Rabi frequencies $\Omega_{c}$ and varied the coupling Rabi frequency $\Omega_{b}$ for each pump Rabi frequency $\Omega_{c}$ and observed the phenomena while changing the coupling Rabi frequency $\Omega_{b}$. For all the three pump Rabi frequencies $\Omega_{c}$, we have observed EIT, EIA and AT splitting depending upon the coupling Rabi frequencies $\Omega_{b}$.

As mentioned earlier, we have scanned the pump beam. We will describe here the observed spectra for the variation of coupling Rabi frequency $\Omega_{b}$ for a fixed pump Rabi frequency $\Omega_{c}$ in detail. In figure \ref{pump_power_1.25mw} we have shown the variation of the probe transmission with the variation of the coupling Rabi frequency $\Omega_{b}$ from $1$ $MHz$ to $43.04$ $MHz$ while the pump Rabi frequency $\Omega_{c}$ was fixed at $17.57$ $MHz$. 

When $\Omega_{b}$ was very less ($1$ $MHz$), EIT was observed (figure \ref{pump_power_1.25mw}(a)). Now if we gradually increase $\Omega_{b}$ keeping $\Omega_{c}$ constant, EIT started to diminish and when $\Omega_{b}\sim \Omega_{c} = 17.57$ $MHz$, EIT was fully transformed into EIA. On further increase of $\Omega_{b}$, EIA started to decrease again. At high $\Omega_{b}$, EIA signal further split and AT splitting occurred. So, just by varying the coupling Rabi frequency $\Omega_{b}$ continuously, the transformation of EIT to EIA to AT was observed as shown in figure \ref{pump_power_1.25mw}(a - i). Here the maximum amplitude of the EIA (figure \ref{pump_power_1.25mw}(e)) is comparable to the initial EIT amplitude (figure \ref{pump_power_1.25mw}(a)).
%Here in this case the maximum amplitude of the EIA peak is greater than that of the EIT as can be seen from figure \ref{pump_power_1.25mw}.

\begin{figure}[h]
	\centering
	\includegraphics[scale=.4]{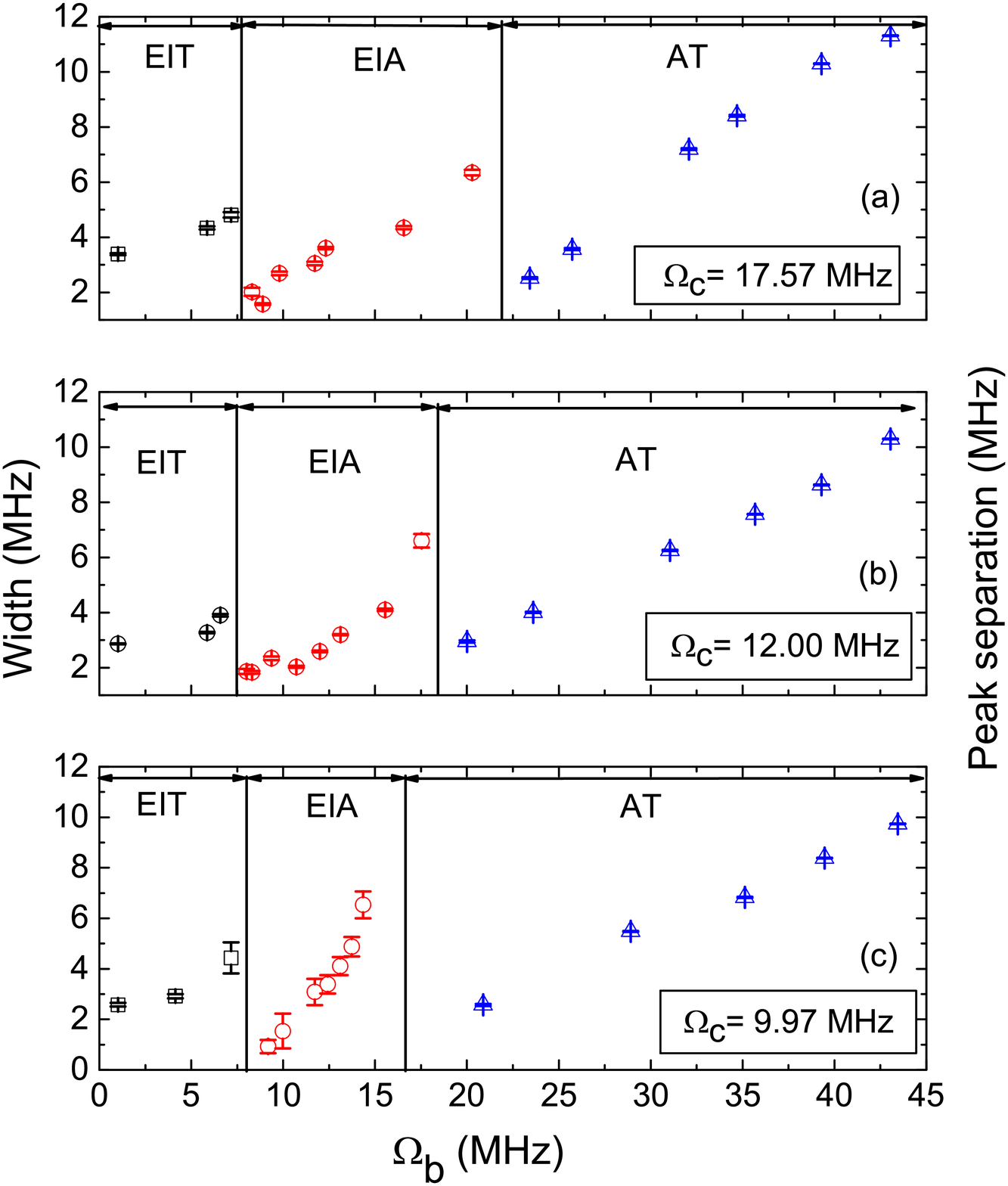}
	
	\caption{Width (EIT, EIA) and peak separations (AT) vs coupling Rabi frequency $\Omega_{b}$ with the pump Rabi frequency $\Omega_{c}$ as a parameter. For the EIT and EIA regions, the plots show the FWHM of the observed peaks and for the AT case, the plots show the peak separations.}
	\label{different regimes}
\end{figure}

The initial EIT line-width was $\sim 3.4$ $MHz$ when $\Omega_{b} = 1$ $MHz$. Till $\Omega_{b} \sim 7.17$ $MHz$ EIT was observed and its line-width changed from $\sim 3.4$ $MHz$ to $\sim 4.81$ $MHz$. From $\Omega_{b}$ $\sim 8.3$ $MHz$ to $20.29$ $MHz$, EIA was observed and its line-width changed from $\sim 2.02$ $MHz$ to $\sim 6.53$ $MHz$. After that the AT regime started. The separation of the AT peaks changed from $\sim 2.52$ $MHz$ to $\sim 11.14$ $MHz$ almost linearly with the increase of $\Omega_{b}$ from $23.42$ $MHz$ to $43.04$ $MHz$ as shown in figure \ref{different regimes}(a).

We have repeated the experiment for two more pump Rabi frequencies i.e., $\Omega_{c} = 12$ $MHz$ and $\Omega_{c} = 9.97$ $MHz$. For both the cases the coupling Rabi frequency $\Omega_{b}$ was varied from $1$ $MHz$ to $43.04$ $MHz$ as earlier. For all the cases, EIT, EIA and AT feature were observed. The main differences were the line-widths and the amplitudes of the observed signals. 

For the case of $\Omega_{c} = 12$ $MHz$, EIT was observed till $\Omega_{b} \sim 6.57$ $MHz$ and its line-width varied from $\sim 2.87$ $MHz$ to $\sim 4$ $MHz$. After that EIA regime started and it was observed till $\Omega_{b} =17.52$ $MHz$. At $\Omega_{b} = 12$ $MHz$, EIT was completely transformed to EIA. Its line-width increased from $\sim 1.83$ $MHz$ to $\sim 6.6$ $MHz$ with increase of $\Omega_{b}$ from $8$ $MHz$ to $17.52$ $MHz$. The separation of AT splitting changed from $\sim 2.95$ $MHz$ to $\sim 10.3$ $MHz$ while $\Omega_{b}$ varies from $20.01$ $MHz$ to $43.04$ $MHz$ (see figure \ref{different regimes}(b)). In this case the maximum amplitude of the EIA was a bit less if compared to the initial EIT amplitude for $\Omega_{c}= 12$ $MHz$.

For $\Omega_{c}= 9.97$ $MHz$, EIT was observed till $\Omega_{b} = 7.17$ $MHz$. Its line-width varied from $\sim 2.5$ $MHz$ to $\sim 4.43$ $MHz$. For the EIA, line-width varied from $\sim 1$ $MHz$ to $\sim 6.5$ $MHz$ with the increase of $\Omega_{b}$  from $7.17$ $MHz$ to $14.35$ $MHz$. After that AT was observed till $\Omega_{b} = 43.04$ $MHz$ and its peak separation increased from $\sim 2.57$ $MHz$ to $\sim 9.74$ $MHz$ (see figure \ref{different regimes}(c)). Here in this case the maximum amplitude of the EIA was quite small as compared to initial EIT amplitude. In the figure \ref{different regimes} (a,b,c) we have approximately shown the three different regions for EIT, EIA and AT splitting when we changed the coupling Rabi frequency $\Omega_{b}$ for each pump Rabi frequency $\Omega_{c}$. We have also shown the FWHM of the EIT and EIA and the separation between the AT peaks as a function of $\Omega_{b}$. It is clear that if we increase the $\Omega_{c}$, the AT splitting started at a higher $\Omega_{b}$ value whereas the EIA started at almost similar $\Omega_{b}$ values for all the cases (figure \ref{different regimes}). Also the FWHM and the AT separation increases for a particular coupling Rabi frequency $\Omega_{b}$ with increase in the pump Rabi frequency $\Omega_{b}$. For the EIT case, the minimum  observed width was $\sim 2.5$ $MHz$ and that for the EIA, it was $\sim 1$ $MHz$.

In the experiment we have also observed the four wave mixing (FWM) peak \cite{Lu:98}  which we have not studied in this article.   

\section{Theoretical model and Discussions}
To uncover the underlying physics behind the experimental observations of the three beam spectroscopy we took two different approaches. Firstly we took the dressed state approach to explain the underlying physical mechanisms for the three different regimes depending only on the coupling Rabi frequency $\Omega_{b}$ variations.  We have also studied the evolutions of the dark state ($\ket{D}$) to the non coupled state $\ket{NC}$. Then we have solved the time dependent Optical-Bloch equations numerically in order to understand the population dynamics in the steady state condition of the different states and to understand the interference contributions in order to attribute the EIT, EIA and AT splitting effects in the system. Further more we have solved the Optical-Bloch equations in steady state condition considering the thermal velocity of the atoms.  We will explain these  approaches in detail in the following sections.
%\newpage

\subsection{Dressed state picture}
\begin{figure}[h]
\centering
\includegraphics[scale=.31]{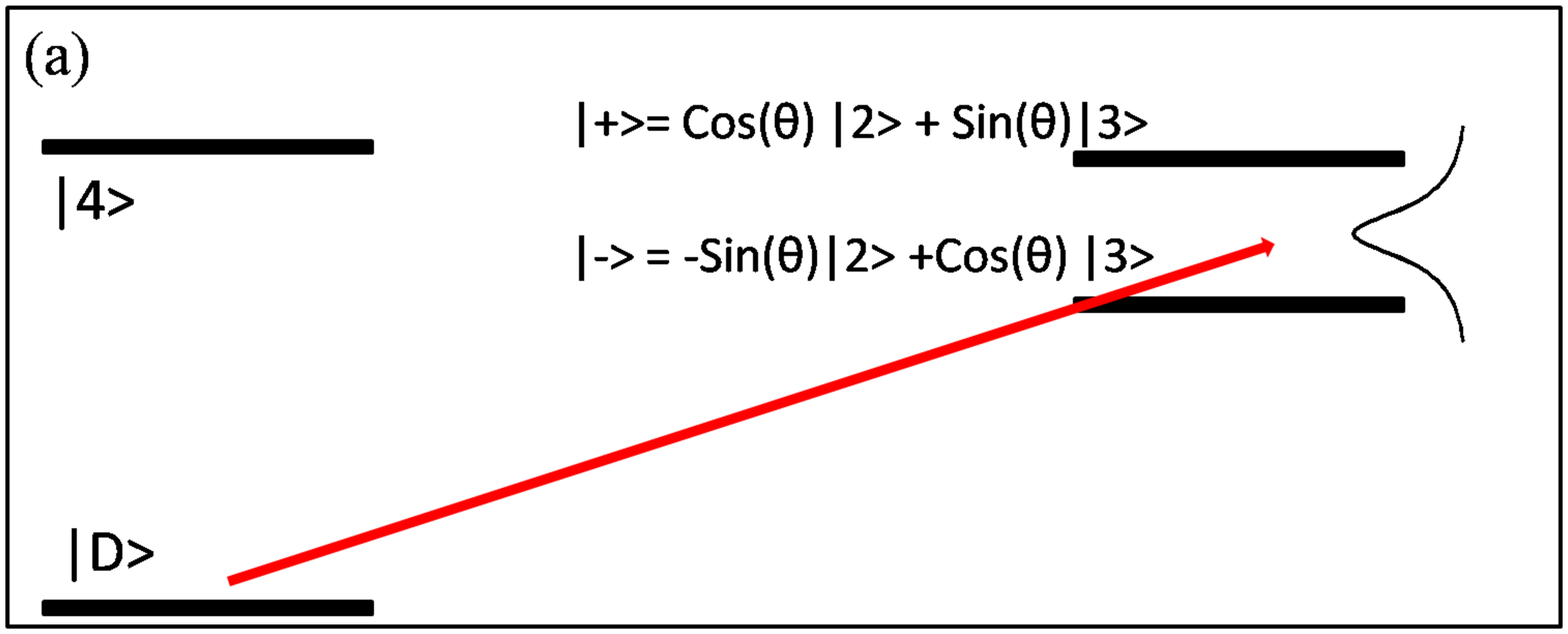}
\includegraphics[scale=.31]{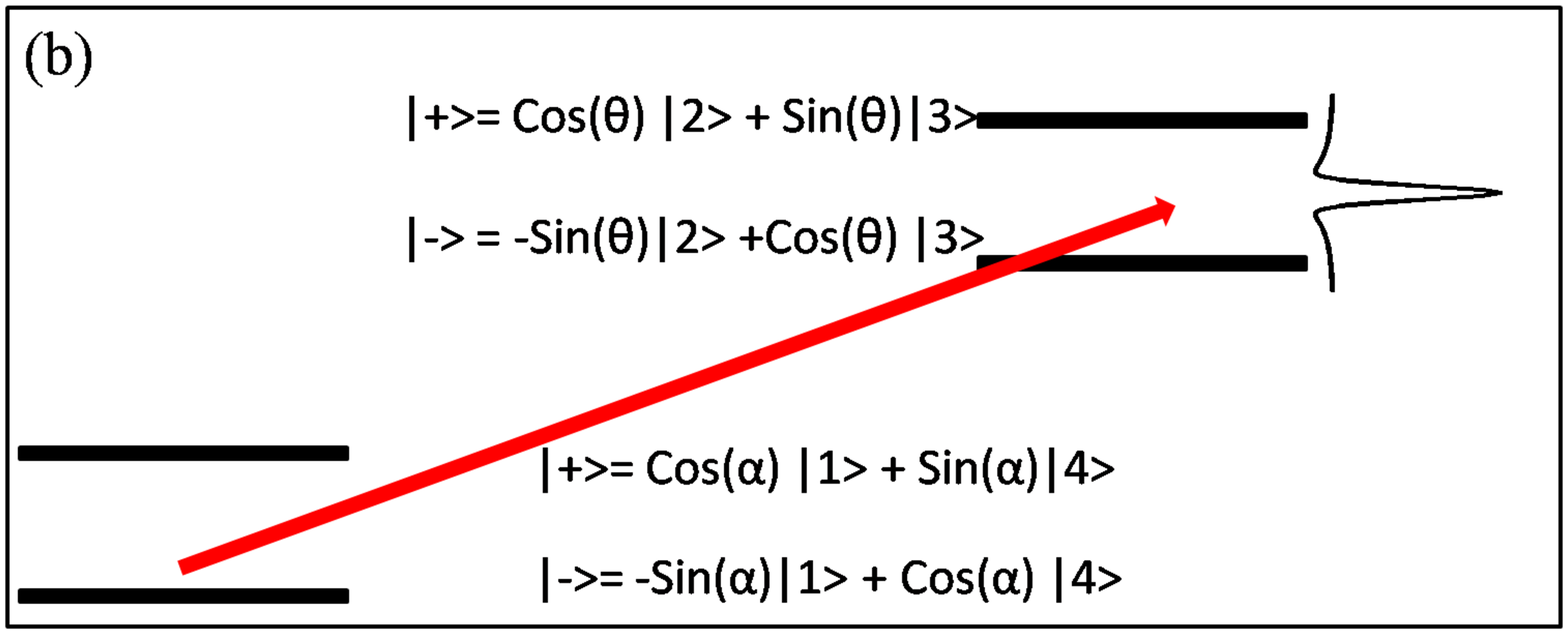}
\includegraphics[scale=.31]{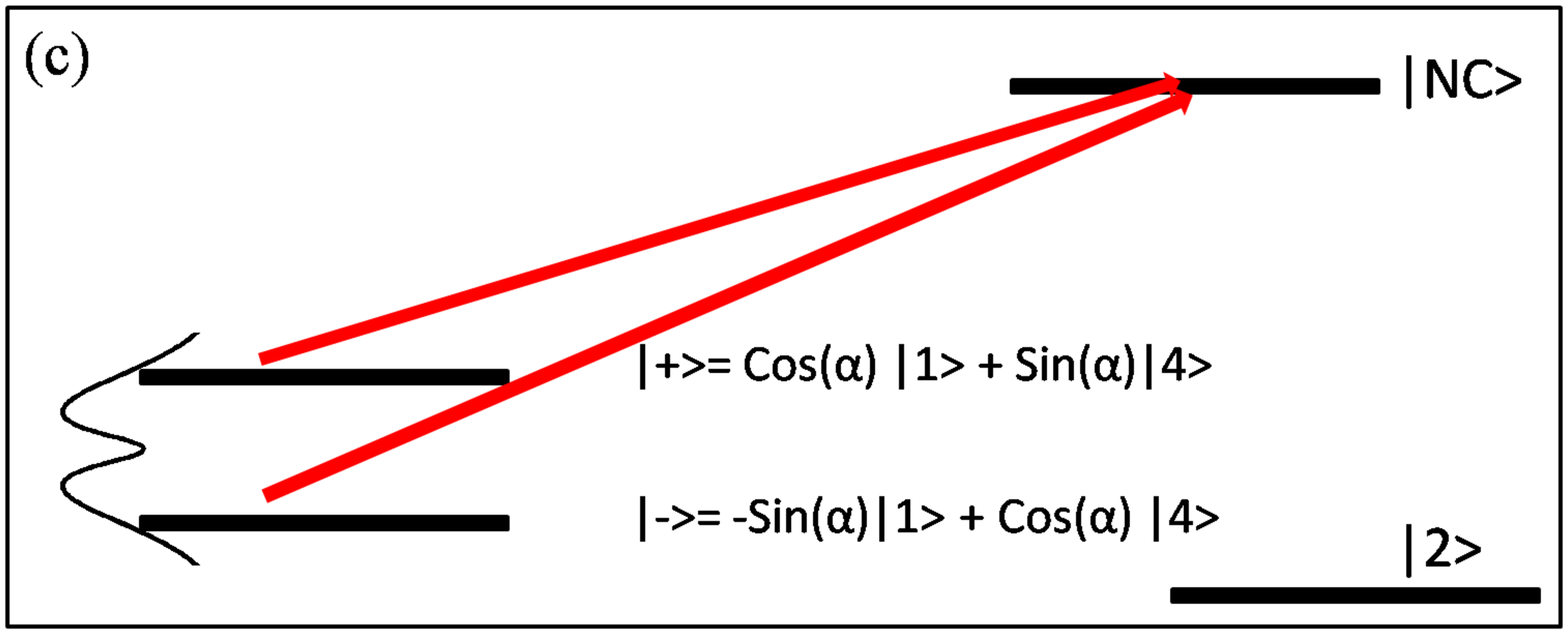}
\caption{Schematic dressed state basis for (a) EIT (b) EIA and (c) AT. Bare states diagram are in the figure \ref{Eperimental setup}(b).}
\label{dressed states}
\end{figure}

In order to understand the physical phenomena acting behind the observation of the three effects, we took the help of the dressed state picture. The total Hamiltonian $\mathcal{H}$ of the system after the rotating wave approximation(RWA), becomes,
\begin{equation}
\mathcal{H}=
\begin{bmatrix}
-\Delta_b + \Delta_p & 0 & -\Omega_p & -\Omega_b\\
0& -\Delta_c &-\Omega_c & 0\\
-\Omega_p& -\Omega_c & \Delta_c - \Delta_p & 0\\
-\Omega_b & 0 & 0& \Delta_b
\end{bmatrix}
\end{equation}
Here, $\Delta_p$, $\Delta_b$, $\Delta_c$ are the detunning of the probe, the coupling and the pump beams respectively. For simplicity of the solution, we will assume that the probe beam is on resonance i.e., $\Delta_p= 0$ and also it is not a dressing field since $\Omega_p<< \Omega_c, \Omega_b$. To explain the different regimes we will consider the partial dressed state concept \cite{nakajima1999}. Now the interaction Hamiltonian can be written as,
\begin{equation}\label{dreesing hamiltonian}
\mathcal{H}=
\begin{bmatrix}
-\Delta_b  & 0 & 0 & -\Omega_b\\
0& -\Delta_c &-\Omega_c & 0\\
0& -\Omega_c & \Delta_c   & 0\\
-\Omega_b & 0 & 0& \Delta_b
\end{bmatrix}
\end{equation}
Now considering the case when $\Omega_b$ is very small  i.e. $\Omega_c$ is the only dressing field, we will partially diagonalize the matrix with respect to the state $\ket{2}$ and $\ket{3}$. 
%we can write the above matrix as follows,
%\begin{equation}
%\begin{array}{cccc}
%0  & 0 & 0 & 0\\
%0& -\Delta_c &-\Omega_c & 0\\
%0& -\Omega_c & \Delta_c   & 0\\
%0 & 0 & 0& 0
%\end{array}
%\end{equation}
If we partially diagonalize the matrix, the eigenvalues become $R_c=\pm \sqrt{\Omega_c^2 -\Delta_c^2}$ and the corresponding rotational matrix becomes,
\begin{equation}
R=
\begin{bmatrix}
1 & 0 & 0 & 0\\
0& \cos{\theta} & \sin{\theta} & 0\\
0&  -\sin{\theta} & \cos{\theta} & 0 \\
0 & 0 & 0& 1 
\end{bmatrix}
\end{equation}
Here, $\cos{\theta} = \dfrac{\Omega_c}{\sqrt{(R_c-\Delta_c)^2 + \Omega_c^2}}$ and $\sin{\theta} = \dfrac{R_c-\Delta_c}{\sqrt{(R_c-\Delta_c)^2 + \Omega_c^2}}$.
So the new basis will be,
\begin{equation}
\vec{U_d}
=
R.\vec{U}
=
\begin{bmatrix}
\Ket{1}\\
\cos{\theta} \ket{2} +\sin{\theta} \ket{3}\\
				 -\sin{\theta} \ket{2} +\cos{\theta} \ket{3}\\
				 \ket{4}
\end{bmatrix}
\end{equation}
Here $\vec{U}= \{\ket{1},\ket{2},\ket{3},\ket{4\}}$ is the bare state vector. It can be observed that $\ket{2}$ and $\ket{3}$ will be dressed and $\ket{1}$ and $\ket{4}$ will be bare states. Here in this case, $\ket{1}$ is the dark state and all the population will be trapped in this state. The fields will destructively interfere and EIT will be observed (see figure \ref{dressed states}(a)).

For the case when both the pump and the coupling beams have comparable Rabi frequencies, both the fields will act as dressing fields. Since the coupling beam is acting on $\ket{1}\rightarrow \ket{4}$, it will dress these two states and the pump beam will dress $\ket{2}$ and $\ket{3}$ states. Since the probe beam will not dress any level, we can decompose the four level system into a two 2-level subsystem i.e. $ \ket{2}, \ket{3}$ and $ \ket{1} ,\ket{4}$. If we diagonalize the Hamiltonian in equation \ref{dreesing hamiltonian}, the eigenvalues will be $ R_b=\pm\sqrt{\Omega_b^2 - \Delta_b^2}$ and $ R_c= \pm \sqrt{\Omega_c^2 - \Delta_c^2}$ and the corresponding transformation matrix will be,
 \begin{equation}\label{EIA rotation matrix}
 R=
 \begin{bmatrix}
 \cos\alpha & 0 & 0 & \sin\alpha\\
 0& \cos{\theta} & \sin{\theta} & 0\\
 0&  -\sin{\theta} & \cos{\theta} & 0 \\
 -\sin\alpha & 0 & 0& \cos\alpha 
 \end{bmatrix}
 \end{equation}
Here $\cos{\alpha} = \dfrac{\Omega_b}{\sqrt{(R_b-\Delta_b)^2 + \Omega_b^2}}$, $\sin{\alpha} = \dfrac{R_b-\Delta_b}{\sqrt{(R_b-\Delta_b)^2 + \Omega_b^2}}$, $\cos{\theta} = \dfrac{\Omega_c}{\sqrt{(R_c-\Delta_c)^2 + \Omega_c^2}}$ and $\sin{\theta} = \dfrac{R_c-\Delta_c}{\sqrt{(R_c-\Delta_c)^2 + \Omega_c^2}}$.
So the new dressed state basis will be
\begin{equation}
\vec{U_d}
=
R.\vec{U}
=
\begin{bmatrix}
\cos{\alpha} \ket{1} +\sin{\alpha} \ket{4}\\
 \cos{\theta} \ket{2} +\sin{\theta} \ket{3}\\
			 -\sin{\theta} \ket{2} +\cos{\theta} \ket{3}\\
			-\sin{\alpha} \ket{1} +\cos{\alpha} \ket{4}
\end{bmatrix}
\end{equation} 

So in this case the dark state $\ket{1}$ which was formed in the EIT case, will be broken due to the coherent interaction of the coupling field $\Omega_b$. Here all the four states will be dressed and they will constructively interfere. As a result the EIA will be observed in the probe transmission (see figure \ref{dressed states}(b)).

For the case when $\Omega_b$ is very high compared to $\Omega_c$ i.e. $\Omega_b>> \Omega_c$, we will again use the partial dressed state concept. Here in this case $\Omega_b$ will be the dressing field. We will partially diagonalized the matrix equation \ref{dreesing hamiltonian} with respect to the states $\ket{1}$ and $\ket{4}$ and  the states $\ket{2}$ and $\ket{3}$ will be bare states.  The eigenvalues become $R_b= \pm \sqrt{\Omega_b^2 -\Delta_b^2}$.
The rotational matrix becomes,
 \begin{equation}\label{AT rotation matrix}
 R=
 \begin{bmatrix}
  \cos\alpha & 0 & 0 & \sin\alpha\\
 0& 1 & 0 & 0\\
 0&  0 & 1 & 0 \\
 -\sin\alpha & 0 & 0& \cos\alpha 
 \end{bmatrix}
 \end{equation} 
Here $\cos{\alpha} = \dfrac{\Omega_b}{\sqrt{(R_b-\Delta_b)^2 + \Omega_b^2}}$ and $\sin{\alpha} = \dfrac{R_b-\Delta_b}{\sqrt{(R_b-\Delta_b)^2 + \Omega_b^2}}$. So the new dressed basis will be,
\begin{equation}
\vec{U_d}= R.\vec{U}=
\begin{bmatrix}
\cos{\alpha} \ket{1} +\sin{\alpha} \ket{4}\\
  \ket{2} \\
 \ket{3}\\
-\sin{\alpha} \ket{1} +\cos{\alpha} \ket{4}
\end{bmatrix}
\end{equation} 

In this case, $\ket{1}$ and $\ket{4}$ will be dressed. Here no population will be trapped  and they will oscillate  in between these two states. Since the probe is acting in between the states $\ket{1}$ and $\ket{3}$, the state $\ket{3}$ will be a non-coupled state $\ket{NC}$. So, no population will be transferred in this state $\ket{3}$. This will be clear in the next section where we have solved the population of each states (figure \ref{numerical simulation}). Here two absorption lines will be observed in the new dressed state and in between these two absorption lines, transparency will be observed resulting in the Aulter-Townes splitting (see figure \ref{dressed states}(c)). In this case since $\Omega_{b}>> \Omega_{c}$, the contribution of the V-type system dominates over the $\Lambda$-type contribution. %Here the Fano interference effects will also be less.

This dressed state analysis helps us to understand the quantum interference phenomena clearly. For the EIT case the states will destructively interfere and the transparency is formed near the resonance. For the EIA case all the four states interfere constructively. In the last case, the non-coupled state $\ket{NC}$ is formed and we get the transparency and the absorption lines due to the superposed states. This is basically AT splitting.

\subsection{ Density matrix solution}

In order to explain the observed phenomena in a more quantitative way, we have solved the optical Bloch equations for a four level atomic system namely, $\mathcal{N}$ level atom. First we have solved the equations of motion in time domain in order to understand the population dynamics and the coherence contributions.  Then we have further solved the equation of motions in the steady state condition for an analytical solution of the medium response. The line-shapes will help us to understand the experimental observations in a more quantitative way.

The master equation \cite{scully} is written as,
\begin{equation}\label{master eq}
\frac{d}{dt}\rho = \frac{-i}{\hbar} [\mathcal{H}, \rho]
\end{equation}

Here $\mathcal{H}$ is the total Hamiltonian of the system and $\rho$ is the density operator. Both $\mathcal{H}$ and $\rho$ are $4\times 4$ matrices. We have added the decay terms phenomenologically.

The equation for the elements can be written as,
\begin{equation}\label{density matrix elements}
\begin{array}{ll}
\dfrac{d\rho_{11}}{dt} &= \Gamma_{31} \rho_{33} + \Gamma_{44} \rho_{44} - \Gamma_{12} \rho_{11} +\Gamma_{21} \rho_{22} - \frac{i}{2} \Omega_{p} \rho_{13} + \frac{i}{2} \Omega^{*}_{p}\rho_{31} - \frac{i}{2} \Omega_{b} \rho_{14} + \frac{i}{2} \Omega^{*}_{b}\rho_{41} \\
\\
\dfrac{d\rho_{22}}{dt} &= \Gamma_{32} \rho_{33}+ \Gamma_{12} \rho_{11} -\Gamma_{21} \rho_{22}- \frac{i}{2} \Omega_{c} \rho_{23} + \frac{i}{2} \Omega^{*}_{c} \rho_{32}\\
\\
\dfrac{d\rho_{33}}{dt} &= -\Gamma_{33}\rho_{33} +  \frac{i}{2} \Omega_{p} \rho_{13} - \frac{i}{2} \Omega^{*}_{p}\rho_{31} + \frac{i}{2} \Omega_{c} \rho_{23} - \frac{i}{2} \Omega^{*}_{c}\rho_{32} \\
\\
\dfrac{d\rho_{44}}{dt} &= -\Gamma_{44} \rho_{44} +  \frac{i}{2} \Omega_{b} \rho_{14} - \frac{i}{2} \Omega^{*}_{b}\rho_{41}\\
\\
\dfrac{d\rho_{14}}{dt} &= -D^{-1}_{14} \rho_{14} + \frac{i}{2} \Omega^{*}_{p} \rho_{34} + \frac{i}{2}\Omega^{*}_{b}(\rho_{44}-\rho_{11})\\
\\
\dfrac{d\rho_{23}}{dt} &= -D^{-1}_{23}\rho_{23} - \frac{i}{2} \Omega^{*}_{p} \rho_{21}+ \frac{i}{2}\Omega_{c}^{*}(\rho_{33}-\rho_{22})\\
\\
\dfrac{d\rho_{21}}{dt} &= -D^{-1}_{21}\rho_{21}-\frac{i}{2}\Omega_{p} \rho_{23}-\frac{i}{2} \Omega_{b} \rho_{24}+ \frac{i}{2} \Omega^{*}_{c} \rho_{31}\\
\\
\dfrac{d\rho_{34}}{dt} &= -D^{-1}_{34}\rho_{34} + \frac{i}{2} \Omega_{p}\rho_{14} + \frac{i}{2} \Omega_{c}\rho_{24}- \frac{i}{2} \Omega^{*}_{b} \rho_{31}\\
\\
\dfrac{d\rho_{24}}{dt} &= -D^{-1}_{24}\rho_{24}+ \frac{i}{2}\Omega^{*}_{c}\rho_{34} - \frac{i}{2}\Omega^{*}_{b}\rho_{21}\\
\\
\dfrac{d\rho_{31}}{dt} &= -D^{-1}_{31}\rho_{31}- \frac{i}{2}\Omega_{p}(\rho_{33}-\rho_{11})- \frac{i}{2}\Omega_{b}\rho_{34} +\frac{i}{2}\Omega_{c}\rho_{21}\\
\end{array}
\end{equation}

Here we have defined $\Gamma_{ij}$ as the spontaneous decay rates of the corresponding levels and $D_{ij}^{-1} = \gamma_{ij} + i
 \Delta_{ij}$. Here $\gamma_{ij}$ are the coherence decay rates from level $i$ to $j$. We have also assumed that the population can decay via collision in between the two ground states $\ket{1}$ and $\ket{2}$ with decay rate $\Gamma_{12}$ and $\Gamma_{21}$. 
Since we have observed the probe transmission, we need to calculate the coherence term $\rho_{31}$.
\subsubsection{Transient solution for density matrix elements}

In order to understand the population dynamics and the coherence effects, first we have solved equation \ref{density matrix elements} in the time domain numerically. Further we have solved the different terms (coherence terms) numerically as a function of time and detunning. In this case we did not consider the thermal averaging for the simplicity of the solution.
 \begin{figure}[h]
 \centering
 \includegraphics[scale=.25]{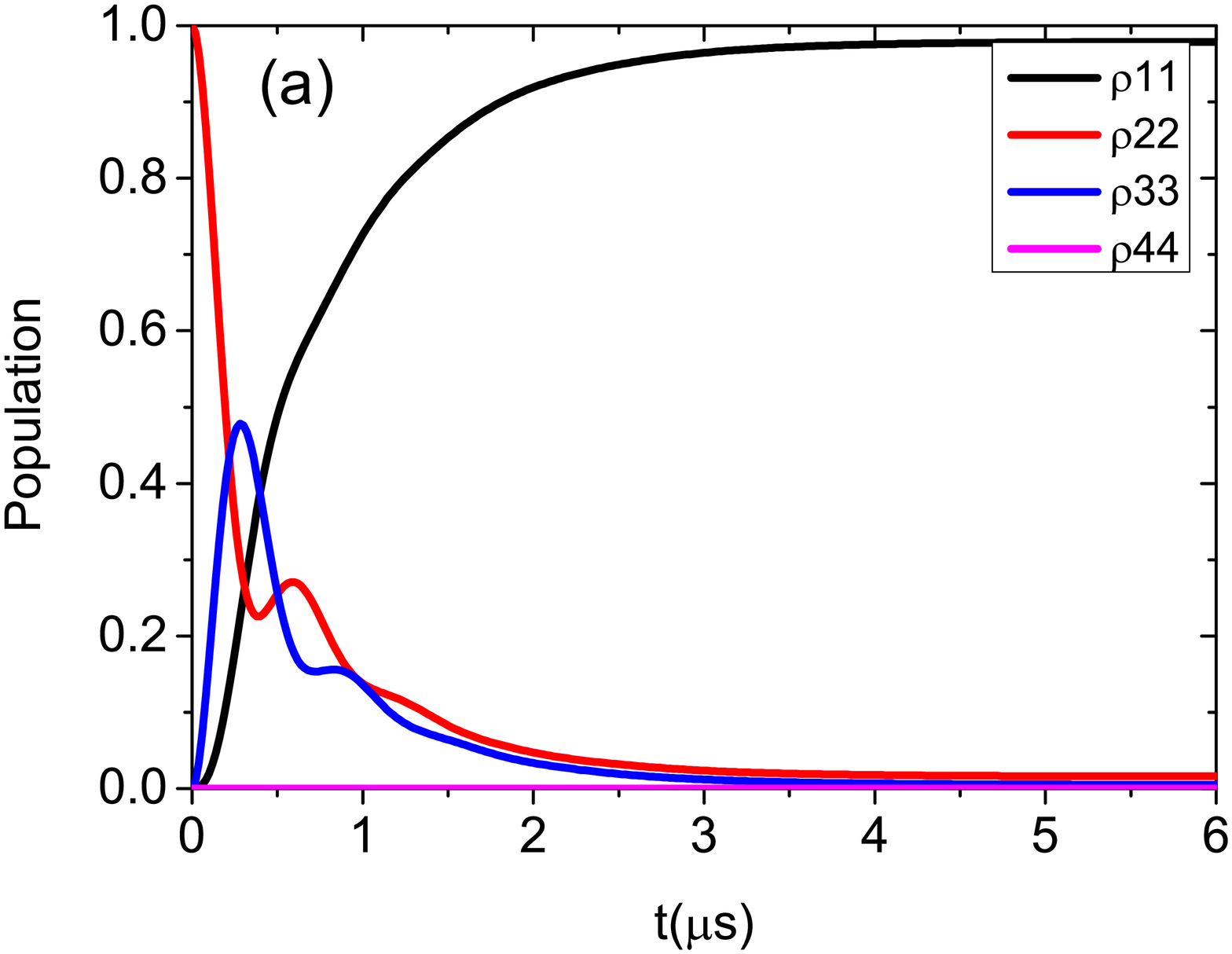}
 \includegraphics[scale=.25]{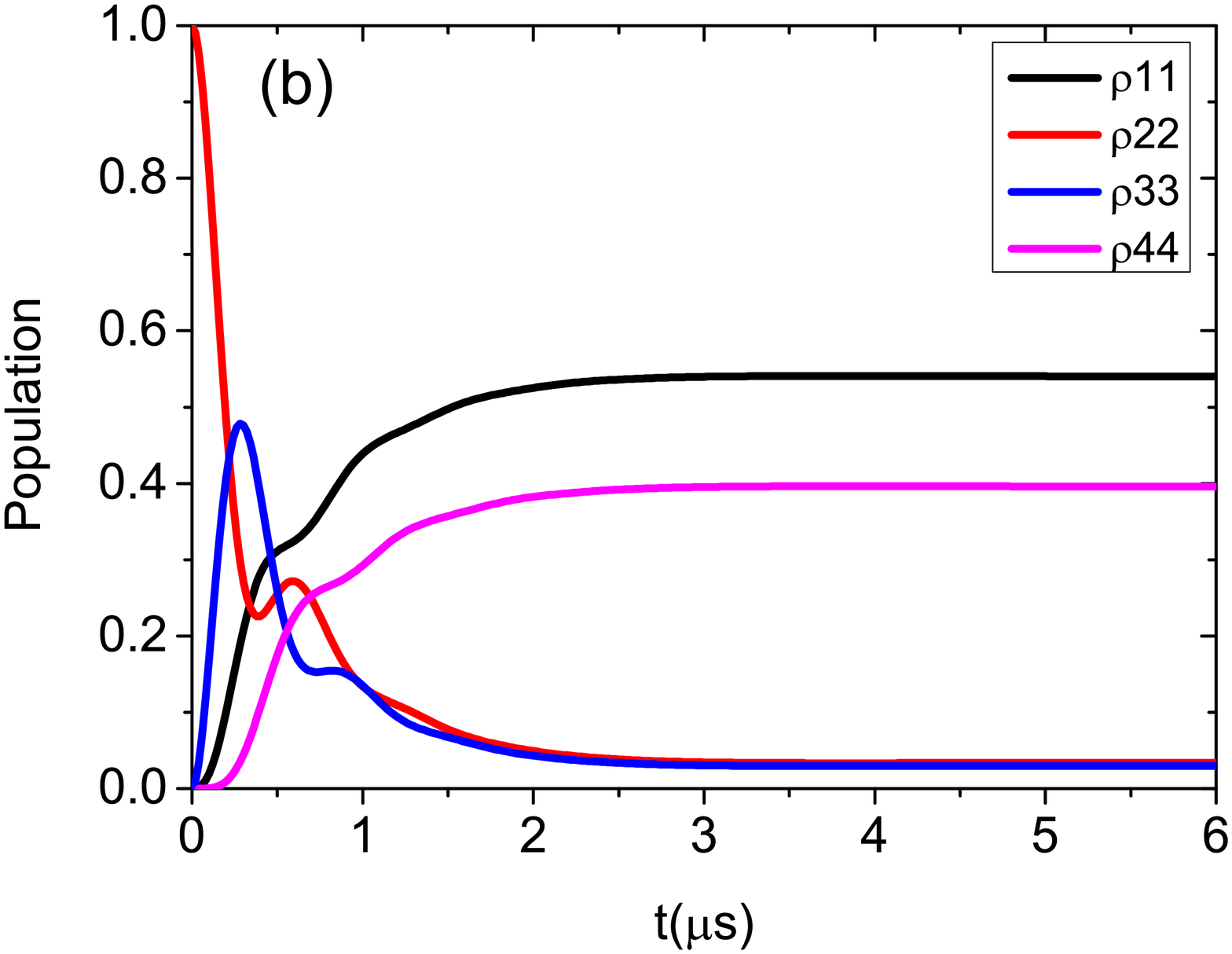}
 \includegraphics[scale=.25]{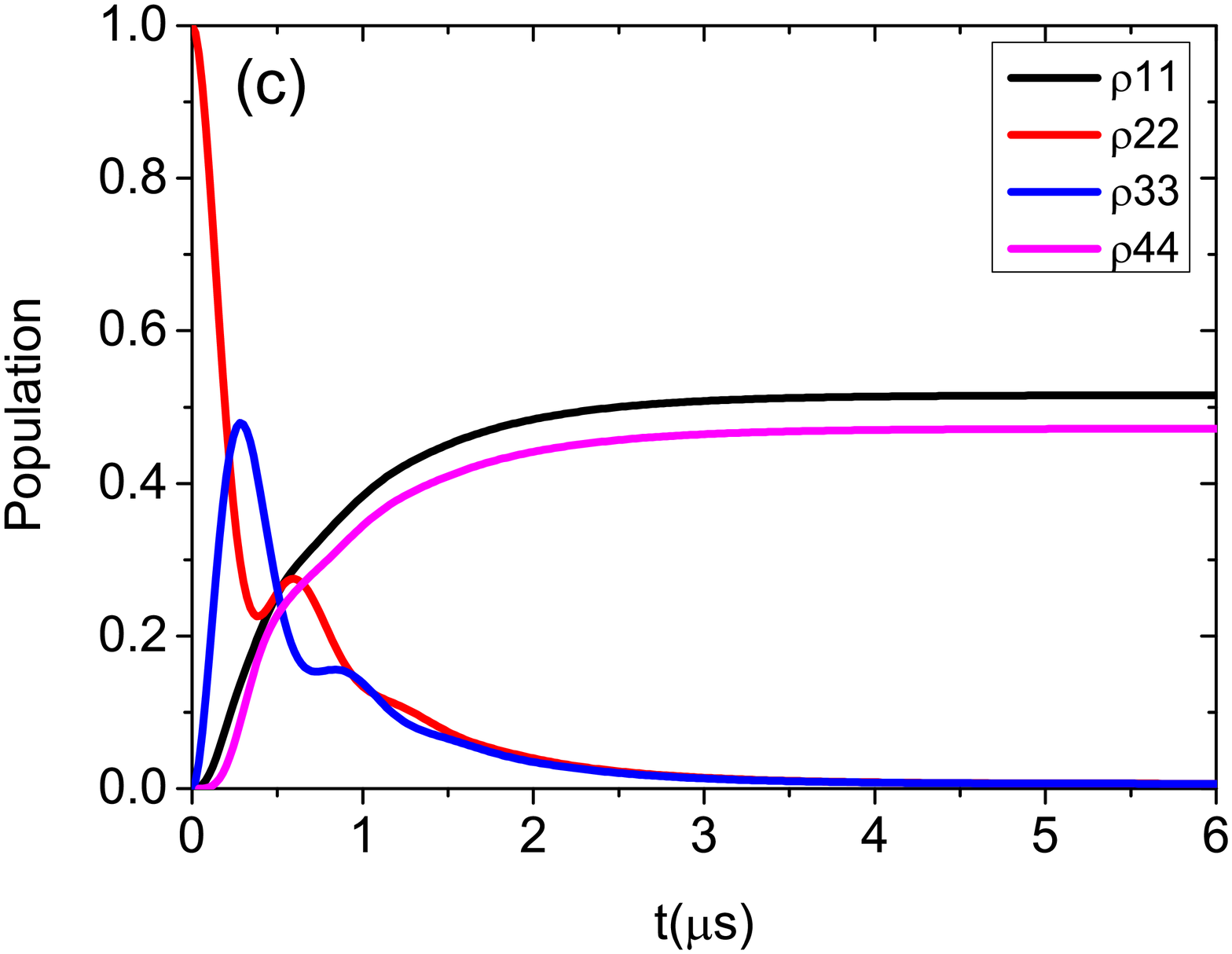}
 
 \caption{ Numerical plots for the populations of the different states (mentioned in the legends). (a) The population distribution for $\Omega_{b}=0$, (b) the population distribution for $\Omega_{b} = 10$ $MHz$ and (c) the population distribution for $\Omega_{b} = 20$ $MHz$. Here the probe Rabi frequency $\Omega_p$ and the pump Rabi frequency $\Omega_c$ are taken to be $1$ $MHz$ and $10$ $MHz$ respectively for all the simulations. }
 \label{numerical simulation}
 \end{figure}
 
In the figure \ref{numerical simulation}(a,b,c) we have plotted the level populations in the three different regimes. Here it was assumed that all the detunnings are zero i.e. all the beams are on resonance. It is observed that all the level populations will be in the steady states after a few Rabi oscillations. We have assumed that initially all the populations are in the state $\ket{2}$.

For the case of EIT, the coupling Rabi frequency $\Omega_{b}$ is zero i.e $\Omega_{b}=0$. Here basically a $\Lambda$-type system is formed consists of the states $\ket{1}$, $\ket{2}$ and $\ket{3}$. Initially after a few Rabi cycles, maximum number of population will be trapped in the dark state $\ket{1}$. These are transferred from the state $\ket{2}$. Due to the collisional decay very few number of atoms will again come to the state $\ket{2}$ from the state $\ket{1}$. Also very few will come in the state $\ket{3}$ since the pump beam is acting in between the states $\ket{2}$ and $\ket{3}$. After the transient time, most of the population will be trapped in the dark state $\ket{1}$ (see figure \ref{numerical simulation}(a)). Due to the  trapping of the populations in the dark state $\Ket{1}$, EIT will be observed in the steady state condition.

For the case where the pump and the coupling Rabi frequency $\Omega_{b}\sim \Omega_{c}$ then the dark state $\Ket{1}$ will be broken and the population will be distributed in the states $\ket{1}$ and $\ket{4}$. In the steady state condition, the population will oscillate in between these two states $\ket{1}$ and $\ket{4}$ when the coupling beam is introduced(see figure \ref{numerical simulation}(b)). The population will be very less in the states $\ket{2}$ and $\ket{3}$ since we have considered the collisional decay between the ground states $\ket{2}$ and $\ket{1}$ . 

\begin{figure}[h]
	\centering
	\includegraphics[scale=.3]{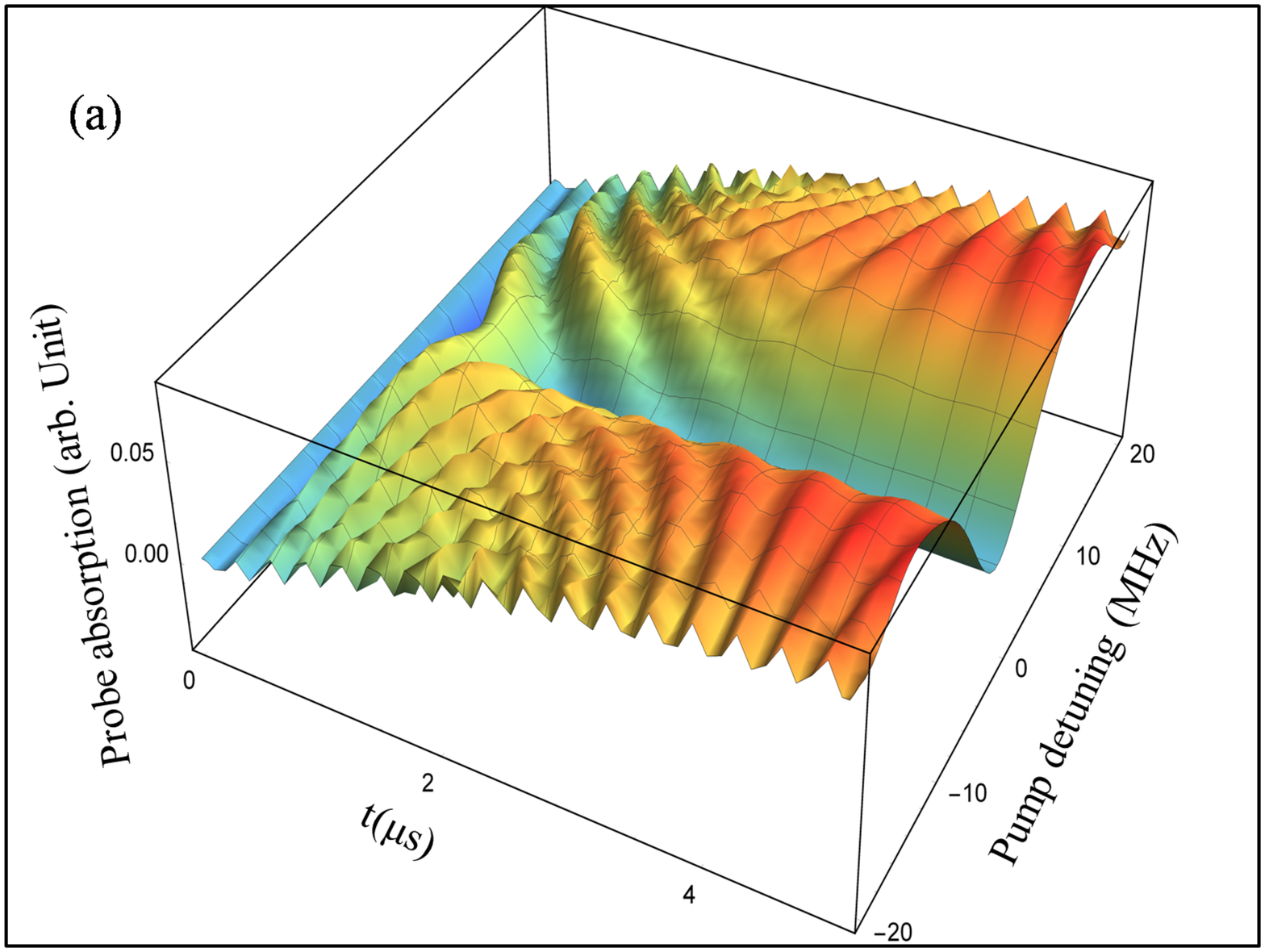}
	\includegraphics[scale=.3]{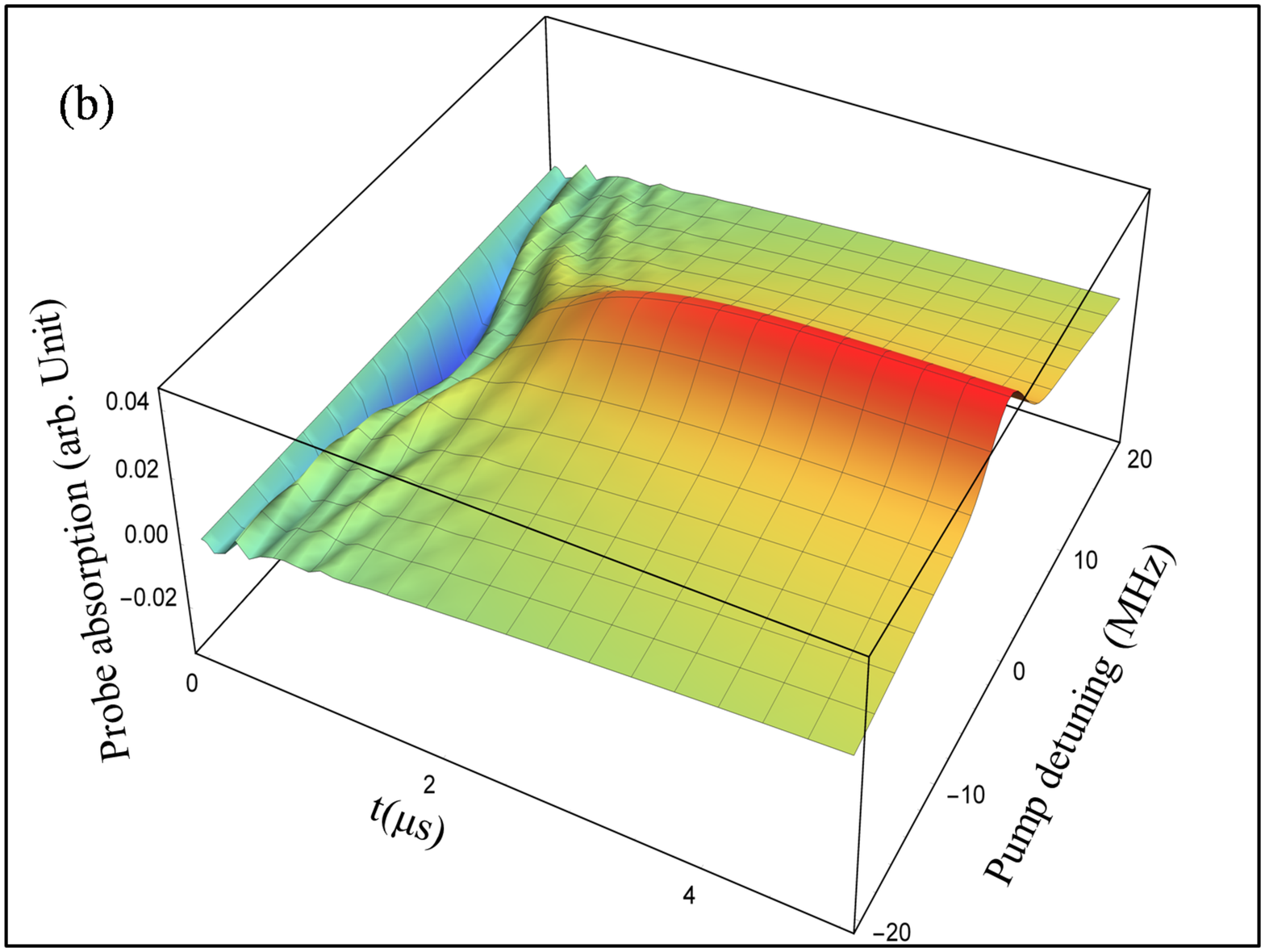}
	\includegraphics[scale=.3]{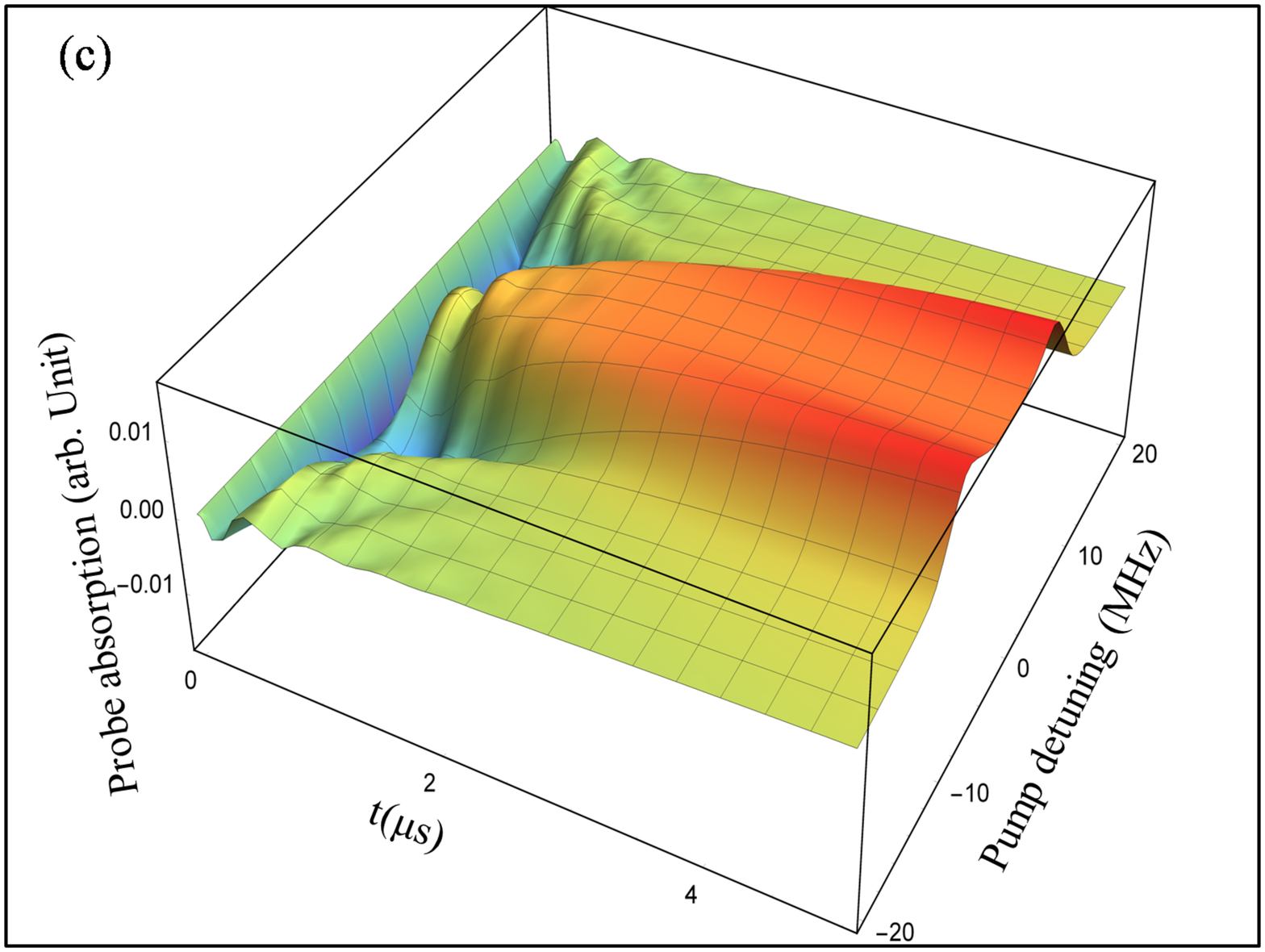}
	
	\caption{ Numerical simulations of the probe coherence term $\rho_{31}$ (probe absorption) as function of time(t) and pump detunning ($\Delta_{c}$). (a) The EIT case $\Omega_{b}=0$, (b) The EIA case $\Omega_{b} = 10$ $MHz$ and (c) the AT case $\Omega_{b} = 20$ $MHz$. Here probe Rabi frequency $\Omega_p$ and pump Rabi frequency $\Omega_c$ are taken to be $1$ $MHz$ and $10$ $MHz$ respectively for all the simulations. }
	\label{numerical simulation detunning}
\end{figure}
For the AT case, when $\Omega_b >> \Omega_c$, we can assume that the V-type contribution is dominating over the $\Lambda$-type system. Since in this case the coupling beam is acting in between the states  $\ket{1}$ and $\ket{4}$, these two states will be dressed and the state $\Ket{3}$ will be a non-coupled state as we have shown earlier. So, the population will be distributed in these two states almost equally. As can be seen from figure \ref{numerical simulation}(c) that the population of  $\ket{1}$ and $\ket{4}$ are almost equal and we will get two absorption lines resulting in the Autler-Townes (AT) splitting.

To understand the coherence effects we have further plotted the probe coherence $Im\{ \rho_{31}\}$ as a function of  pump detunning ($\Delta_{c}$) and time$(t)$. These solutions are obtained by solving the equations \ref{density matrix elements} numerically in time (t) and detunning $(\Delta_{c})$ domain.  The profiles along the detunning axis signifies the probe transmission which we have observed experimentally in the steady state condition.
 
It can be observed from figure \ref{numerical simulation detunning} (a,b,c) that after a few Rabi cycles, the system reached the steady state and we observed the steady line-shapes along the detunning axis. The results of the steady line-shapes will be solved in the steady state condition in the next section. In the 1st case it is observed that when $\Omega_{b}= 0$, EIT is observed along the detunning axis (figure \ref{numerical simulation detunning}(a)). When $\Omega_{b} = \Omega_c$, EIA is observed again along the detunning axis as shown infigure \ref{numerical simulation detunning}(b). On increasing the coupling Rabi frequency $\Omega_b$, EIT is transformed into EIA.  For the third case when  $\Omega_{b} >> \Omega_c$, the AT spliting occurs. In figure \ref{numerical simulation detunning}(c) two absorption lines can be seen and in between them the transparency is observed.

\subsubsection{Steady state solution considering thermal averaging}

Here we have solved the Optical-Bloch equations (equation \ref{density matrix elements}) in the steady state. Here we have considered the velocity of the atoms since in the experiment, the velocity of the atoms comes into the picture. The coherence term $\rho_{31}$ will give us the probe transmission. In the steady state condition $\rho_{31}$ becomes,
 \begin{equation}\label{probe coherence}
 \rho_{31}= \dfrac{-\frac{i}{2} \Omega_{p} \left(\rho_{33}^{0}-\rho_{11}^{0}\right) + \frac{\Omega_{p}\Omega_{b}}{4}D_{34}\left[1-\frac{\Omega_{c}^{2}}{4}A_{24}(D_{34}+ D_{21})\right]\rho^{0}_{14} + \frac{\Omega_{p}\Omega_{c}}{4}D_{21}\left[1-\frac{\Omega_{b}^{2}}{4}A_{24}(D_{34}+ D_{21})\right]\rho^{0}_{23}}{A^{-1}_{31} - \frac{\Omega_{b}^2 \Omega_{c}^2}{16}(D_{34} +D_{21})^{2} A_{24}}
 \end{equation}

Here we have defined $A_{31}$, $A_{24}$ as follows and also $\rho_{14}$, $\rho_{23}$ to be,
\begin{equation}
\begin{array}{ll}
A_{31}^{-1} &= D_{31}^{-1} + \frac{\Omega_{c}^{2}}{4} D_{21} + \frac{\Omega_{b}^{2}}{4} D_{34}\\
\\
A_{24}^{-1} &= D_{24}^{-1} + \frac{\Omega_{c}^{2}}{4} D_{34} + \frac{\Omega_{b}^{2}}{4} D_{21}\\
\\
\rho_{14}^{0} &=\frac{i}{2}D_{14}\Omega_{b}^{*} \left(\rho^{0}_{44}-\rho^{0}_{11}\right)\\
\\
\rho_{23}^{0} &=\frac{i}{2}D_{23}\Omega_{c}^{*} \left(\rho^{0}_{33}-\rho^{0}_{22}\right)
\end{array}
\end{equation}

Now the zeroth order population can be calculated by the following relations. We will solve for the $\rho^{0}_{22}$ term and from that all the population can be calculated.
\begin{equation} \label{steady populations}
\begin{array}{ll}
\rho^{0}_{22} &=\\
 &\frac{\Gamma_{12}}{2}\times\frac{[\Omega_{c}^2 + \Gamma^2 + 4(\Delta_{c}+kv)^2 ][\Omega_{b}^2 + \Gamma^2 + 4(\Delta_{b}+kv)^2 ]}{\frac{\Gamma_{21}}{2} [\Omega_{c}^2 + \Gamma^2 + 4(\Delta_{c}+kv)^2 ][2\Omega_{b}^2 + \Gamma^2 + 4(\Delta_{b}+kv)^2 ]+ \frac{\Gamma_{12}}{2} [2\Omega_{c}^2 + \Gamma^2 + 4(\Delta_{c}+kv)^2 ][\Omega_{b}^2 + \Gamma^2 + 4(\Delta_{b}+kv)^2 ] + \frac{\Omega_{c}^2\Gamma}{4} [2\Omega_{b}^2 + \Gamma^2 + 4(\Delta_{b}+kv)^2 ]}\\
\\

\rho^{0}_{33} &= \dfrac{\Omega_{c}^2 }{\Omega_{c}^2  +  (\Gamma^2 + 4(\Delta_{c} +kv)^2)}\times \rho_{22}^{0} \\
\\
\rho^{0}_{11}&= \dfrac{\Omega_{b}^2 + \Gamma^2 + 4 (\Delta_{b} + kv)^2}{2\Omega_{b}^2 + \Gamma^2 + 4 (\Delta_{b} + kv)^2}- \dfrac{2\Omega_{c}^2 + \Gamma^2 + 4 (\Delta_{c} + kv)^2}{\Omega_{c}^2 + \Gamma^2 + 4 (\Delta_{c} + kv)^2}\times \dfrac{\Omega_{b}^2 + \Gamma^2 + 4 (\Delta_{b} + kv)^2}{2\Omega_{b}^2 + \Gamma^2 + 4 (\Delta_{b} + kv)^2}\times \rho^{0}_{22} 
\\
\rho^{0}_{44} &= \dfrac{\Omega_{b}^2 }{\Omega_{b}^2 +  (\Gamma^2 + 4(\Delta_{b}+kv)^2)}\times\rho^{0}_{11}
\end{array}
\end{equation}

In these calculations we have considered collisional decay between the ground states i.e. $\Gamma_{12}= \Gamma_{21}\neq 0$. For this condition, the population will be distributed into all the four levels. The population will be very less in the states $\ket{2}$ and $\ket{3}$ as $\Gamma_{12}$ and $\Gamma_{21}$ are very small $(\sim 300$ $KHz)$. Most of the populations will be in the states $\ket{1}$ and $\ket{4}$ which can be seen from the above equations \ref{steady populations}. We have already observed this argument in the previous section from the population calculations in time domain (figure \ref{numerical simulation}) . But if $\Gamma_{12}= \Gamma_{21}= 0$, then population will be only in the states $\ket{1}$ and $\ket{4}$. In the above calculations $\Omega_{p}$, $\Omega_{c}$,  and $\Omega_{b}$ are assumed to be real.

%In the calculation of the coherence term we have assumed that initially all the populations are in the ground state $\ket{2}$. But after few Rabi cycle most of  populations will be transferred to the state $\ket{1}$ which we have already shown in the earlier section. Since we have considered the collisional decay the populations are distributed in all the levels. When the coupling beam is introduced the population will oscillate in between states $\ket{1}$ and $\ket{4}$ as can be seen from figure \ref{numerical simulation}. Very few atoms will be in the states $\ket{2}$ and $\ket{3}$. 

In the calculation of the coherence term $\rho_{31}$ in equation \ref{probe coherence} we have assumed that the transitions $\ket{2}\rightarrow\ket{4}$, $\ket{2}\rightarrow\ket{1}$, and $\ket{3}\rightarrow\ket{4}$ are not dipole allowed transitions. This equation \ref{probe coherence} is actually the analytical solution of the line shape of the numerically solved results of figure \ref{numerical simulation detunning} along the detunning axis for zero velocity group of atoms only when the system have reached the steady state.\\ 

Now in order to calculate the total probe transmission or the susceptibility $\chi$ of the medium we need to consider the thermal velocity of atoms. The atoms obey Maxwell-Boltzmann (M-B) velocity distribution. Considering all the possible velocities the susceptibility becomes,
\begin{equation}\label{chi}
\chi  = \dfrac{\mu}{\epsilon_{0} E_p}\int_{-\infty}^{\infty} N(k v) \rho_{31} d(k v)
\end{equation}

The distribution for a velocity $v$ can be written as,
\begin{equation}
N(kv) = \frac{N_0}{\sqrt{\pi k^2 u^2}} e^{-(k v)^2/(ku)^2}
\end{equation}

\begin{figure}[h]
\centering
\includegraphics[scale=.25]{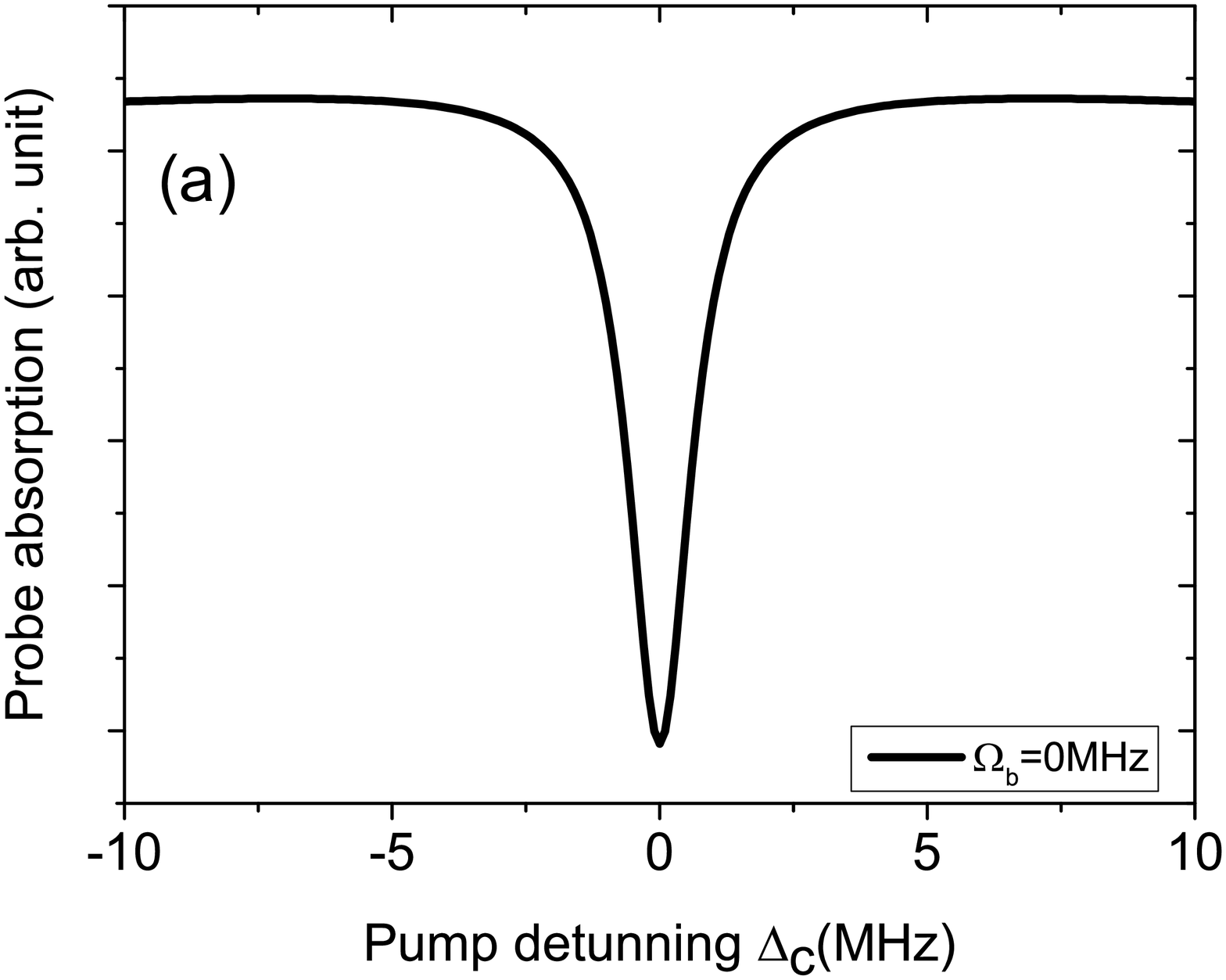}
\includegraphics[scale=.25]{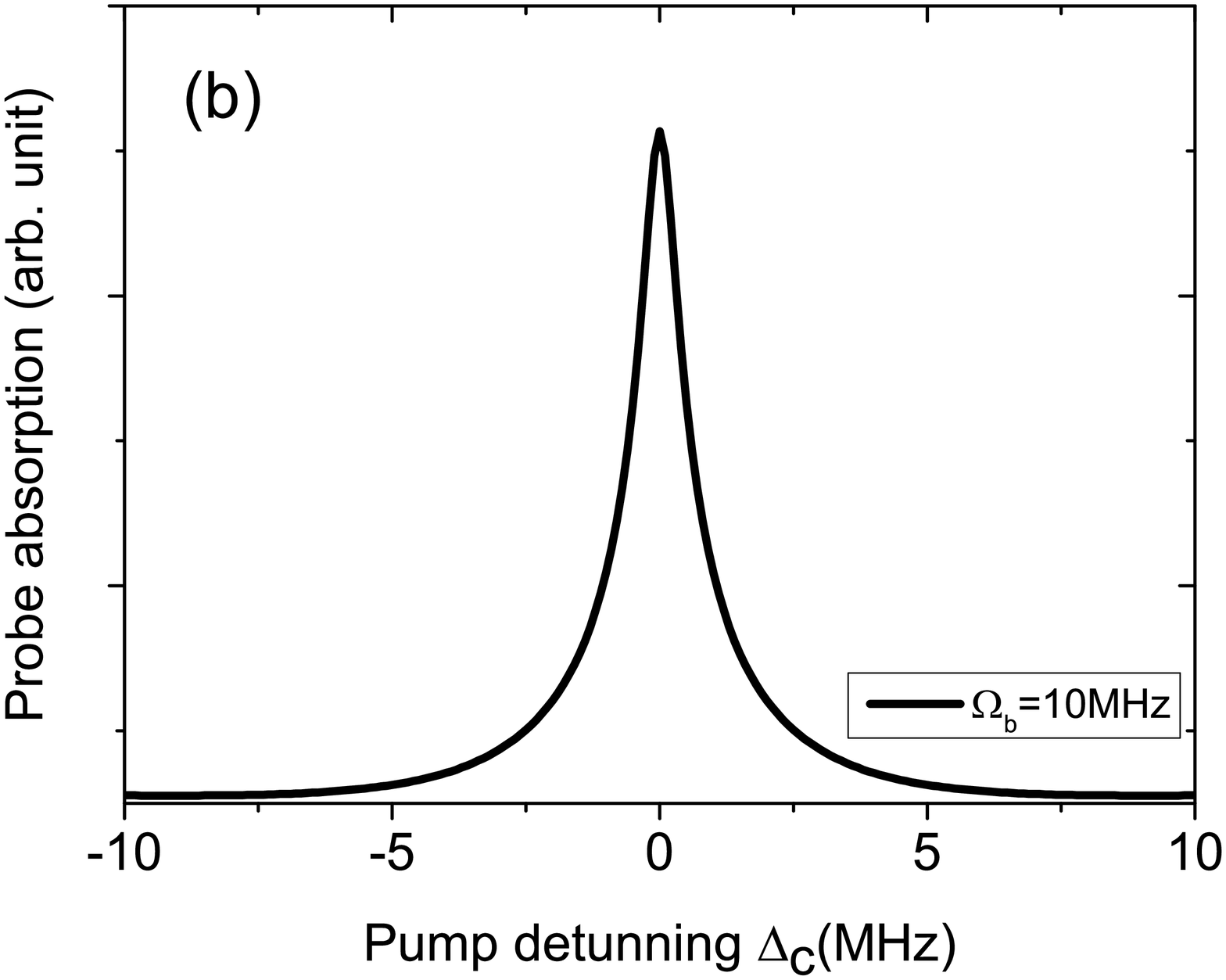}
\includegraphics[scale=.25]{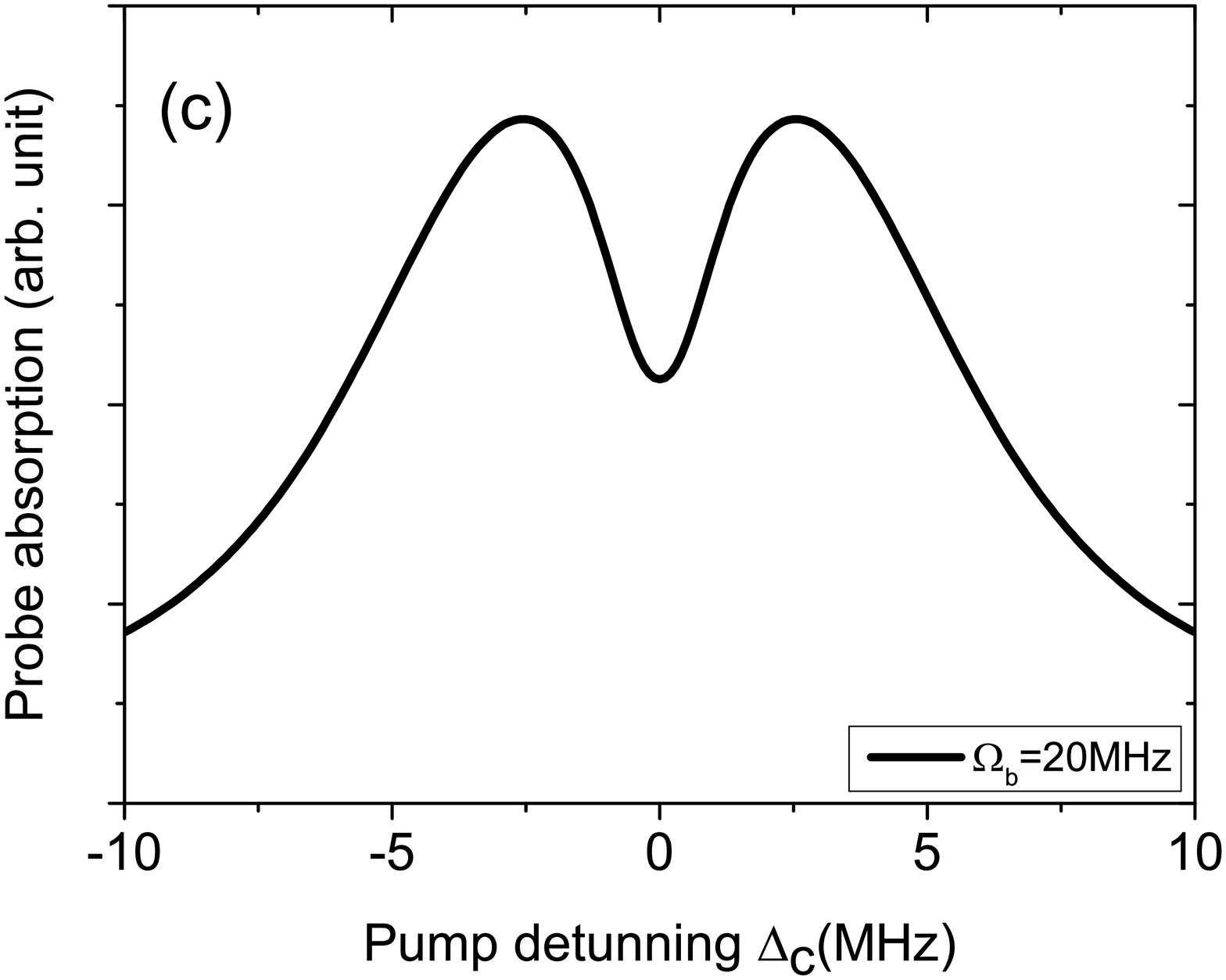}

\caption{ Theoretical simulation of the probe coherence $\rho_{31}$ as a functions of pump detunning $\Delta_{c}$ considering the M-B distribution (a) The EIT case when $\Omega_{b} = 0$ $MHz$, (b) the EIA case when $\Omega_{b} = 10$ $MHz$ and (c) the AT case when $\Omega_{b} = 20$ $MHz$. Here the probe Rabi frequency $\Omega_p = 1$ $MHz$ and the pump Rabi frequency $\Omega_c = 10$ $MHz$ are taken for all the simulations.}
\label{theoretical simulation}
\end{figure}

In the figure \ref{theoretical simulation} (a,b,c) we have plotted the numerical results of velocity averaging with the pump detunning for different coupling Rabi frequencies $\Omega_b$ as we did in the experiment. Since we are scanning the pump frequency, the Doppler background part is suppressed and we are only getting the EIT, EIA and AT features. These figures are similar to the detunning axis of figure \ref{numerical simulation detunning}. But here the velocity of the atoms are taken into account. Due to this, the Doppler narrowing in the line-widths of the EIT, EIA and AT are observed. Similar to the earlier cases, when $\Omega_{b}=0$ but $\Omega_{c}\neq 0$, EIT is observed. When $\Omega_{b}=\Omega_{c}$, EIA is observed and $\Omega_{b}>>\Omega_{c}$ results in the AT effect.\\

Now the above observations can be explained in the following way. Initially when the coupling channel was not opened it was basically a $\Lambda$-type system. All the population was trapped in the dark state $\ket{1}$ due to the two photon resonance and EIT was observed. When the coupling channel was opened and we were increasing the coupling Rabi frequency gradually basically, we were breaking the dark state of the system. As a result the $\Lambda$-type contribution started to decrease and it behaved more likely as the $\mathcal{N}$-type system. The dark state $\ket{1}$ got modified into two new superposition states i.e., $\cos{\alpha} \ket{1} +\sin{\alpha} \ket{4}$ and $-\sin{\alpha} \ket{1} +\cos{\alpha} \ket{4}$ as shown earlier.

When the pump and the coupling Rabi frequencies were comparable the EIT was fully transformed into EIA. This was a three photon resonance feature where all the states constructively interfere. This observed EIA was due to the transfer of coherence (TOC) \cite{goren,Taichenachev99}. The pump and the coupling beams had the same polarization but the probe had different polarization. The excited states coherence $\rho_{34}$ has a non zero contribution in the ground state coherence $\rho_{21}$. Since the transition $\ket{4}\rightarrow \ket{1}$ is a closed transition the TOC ($\rho_{34}$) is non zero here. This observed EIA was slightly different from what was predicted by Goren et. al \cite{goren}. They have considered the system to be in Zeeman sub-levels but here we observed the phenomena in the hyperfine levels. The conditions for observing the EIA was also not strictly followed. In our case, $\ket{3}\rightarrow \ket{2}$ is an open hyperfine transition but $\ket{4}\rightarrow \ket{1}$ is a closed hyperfine transition.

We have considered small collisional decays $\Gamma_{12}$ and $\Gamma_{21}$ between the ground states theoretically. Even then we observed the formation of EIA but its amplitude is little less than that compared to the zero collisional decay. Therefore these collisions are decreasing the effects of TOC in this experiment, although its effect is small.

Now with further increase of the coupling Rabi frequency $\Omega_b$, the $\mathcal{N}$-type contribution started to decrease and more likely the $V$-type contribution started to dominate. Here the EIA was further splitted which was basically AT splitting. The population of the $\ket{1} $ and $\ket{4}$ states became almost equal and we observed AT splitting. We found the peak separation to be increasing with the increase of the coupling Rabi frequency $\Omega_b$ (see figure \ref{different regimes} (a,b,c)) which is also a signature of AT.

In our experiment the transition $\ket{2}\rightarrow \ket{4}$ i.e $F= 2\rightarrow F'=4$ is not dipole allowed transition. So the spontaneous decay $\Gamma_{24}$ is zero.  Instead, if the transition $\ket{4}\rightarrow \ket{2}$ was allowed, then the EIA would not have been observed which we checked experimentally in the case of $^{87}Rb$-$D_2$ transition. Our theoretical simulation also supports this statement. 

EIT to AT transformations were studied in several systems \cite{barry2011} and a crossover \cite{Zhu2013,Tan:14} exists between them. But in our case this crossover was replaced by the EIA interference contribution which can easily be studied using the above formulations.

\section{Conclusion}
In this article we have shown how the atomic states can be engineered. How the response of the atomic system can be modulated from highly transparent to highly absorbing one by changing only the coupling Rabi frequency $\Omega_b$. Further this highly absorbing medium can be made to split into two absorption lines resulting in a transparency in between them. We have studied the interplay between the coherent phenomena EIT, EIA and the AT splitting. The theoretical models helped us to uncover the underlying physics behind them. In the presence of the third beam (coupling beam), the coherent interaction led to the splitting of  the dark states to another superposition states resulting into sharp spectral features.\\ 
The analytical solution of the medium coherence term gave more insight into how in the presence of the coupling beam gave rise to different features. The partial dressed state concept helped us to decouple the system into two simple 2-level systems. This helped us to understand the physical mechanisms responsible for the interplay in a very simple manner and how the dressing contribution changes depending upon the coupling and the pump Rabi frequencies.

Apart from understanding some fundamental features, the system can be used in optical switching applications since the medium can be tunned from highly transparent to highly absorbing medium \cite{Sheng:11}. All the line-widths that were observed are sub-natural line-widths, so this can be used for precision experiments. The transformation of subluminal light to superluminal light propagation can also be studied in this system. This system may be useful for making white light cavity since in the AT regime it can have anomalous dispersion in the transparent background \cite{WICHT1997}.

\section{Acknowledgement}
B.C.D., A.D. and S.D. acknowledge the financial support received from the Department of Atomic Energy, Government of India (Grant No. 12-R\&D-SIN-5.02-0102). S.C. thanks the Department of Science and Technology (DST), Government of India for the project grant under women scientist scheme WOS-A (Sanction No. SR/WOS-A/PM-1040/2014(G)).

%\bibliography{bibliography.bib}
\end{document}